\documentclass[journal]{IEEEtran}

\usepackage{cite}
\usepackage{graphicx}
\usepackage{psfrag}
\usepackage{subfigure}
\usepackage{url}
\usepackage{stfloats}
\usepackage{amsmath}
\usepackage{array}
\usepackage{amssymb}
\usepackage{makecell}
\usepackage{stfloats}
\usepackage{booktabs}
\usepackage{threeparttable}
\usepackage{multirow}

\allowdisplaybreaks[4]

\ifCLASSINFOpdf

\else

\fi

\begin{document} \title{A Comparative Study of Unipolar OFDM Schemes in Gaussian Optical Intensity Channel}

\author{Jing~Zhou, \emph{Member, IEEE},
        and~Wenyi~Zhang, \emph{Senior Member, IEEE}

\thanks{This work was supported in part by the Key Research Program of Frontier Sciences of CAS under Grant QYZDY-SSW-JSC003, by the National Natural Science Foundation of China under Grants 61379003 and 61722114, and by the Fundamental Research Funds for the Central Universities under Grants WK2100060020 and WK3500000003. This paper was presented in part at WCSP 2017 \cite{WCSP}.

The authors are with the Key Laboratory of Wireless-Optical Communications, Chinese Academy of Sciences, and with the Department
of Electronic Engineering and Information Science, University of Science and Technology of China, Hefei, China (e-mail: jzee@ustc.edu.cn; wenyizha@ustc.edu.cn).}}

\maketitle

\begin{abstract}

We study the information rates of unipolar orthogonal frequency division multiplexing (OFDM) in discrete-time optical intensity channels (OIC) with Gaussian noise under average optical power constraint. Several single-, double-, and multi-component unipolar OFDM schemes are considered under the assumption that independent and identically distributed (i.i.d.) Gaussian or complex Gaussian codebook ensemble and nearest neighbor decoding (minimum Euclidean distance decoding) are used. We obtain an array of information rate result. These results validate existing signal-to-noise-and-distortion-ratio (SNDR) based rate analysis, establish the equivalence of information rates of certain schemes, and demonstrate the evident benefits of using component-multiplexing at high signal-to-noise-ratio (SNR). For double- and multi-component schemes, the component power allocation strategies that maximize the information rates are investigated. In particular, by utilizing a power allocation strategy, we prove that several multi-component schemes approach the high SNR capacity of the discrete-time Gaussian OIC under average power constraint to within 0.07 bits.
 \end{abstract}


\section{Introduction}
\subsection{Background, Related Work, and Motivation}

Orthogonal frequency division multiplexing (OFDM) has been widely used in wireline and wireless communications. In optical communications including optical fiber and optical wireless communications (OWC), OFDM is a promising technique \cite{Armstrong}. Many optical systems are based on intensity modulation and direct detection (IM/DD), which carries information by varying optical intensity (the optical power transferred per unit area). Therefore in IM/DD systems a primary concern is designing nonnegative (unipolar) transmit signals. For OFDM this task is challenging because it is difficult to ensure the unipolar property by simply constraining the discrete frequency domain input symbols.

The direct current (DC) offset OFDM (DCO-OFDM) \cite{B94,CK96} and the asymmetrically clipped OFDM (ACO-OFDM) \cite{AL06} are two well-known unipolar OFDM schemes. The pulse-amplitude-modulated discrete multitone modulation (PAM-DMT) \cite{LRBK09} and the Flip-OFDM \cite{Yong07,FHV11,TSH12} are also widely investigated in literature.
Unfortunately, the unipolar constraint inevitably reduces spectral efficiency. Among the above four schemes, the DCO-OFDM loses efficiency in terms of optical power due to the DC bias added, while the other three lose efficiency in terms of bandwidth/degrees of freedom (DoF)\footnote{The DoF efficiency, which is a dimensionless measure of bandwidth efficiency, stands for the number of independent symbols transmitted per channel use in time domain, divided by the maximum symbol rate of the channel, which is one real symbol ($\frac{1}{2}$ complex symbol) per channel use for optical intensity channel with single input and single output. Note that the DoF efficiency determines the pre-log factor of the information rate achieved.}  due to their constraints on the discrete frequency domain input (see Table I for details).

\begin{table*}[tbp]
\renewcommand{\cellset}{\renewcommand{\arraystretch}{1.1}}
\centering
\caption{Unipolar OFDM schemes Considered in This Paper}
\begin{threeparttable}
\scalebox{1}{
\begin{tabular}{c|c|l|l|l|l|l|l|l}
\hline
\multicolumn{2}{c|}{}&\makecell[cc]{\textbf{Scheme}}&\makecell[cc]{\textbf{Basic Idea} \\\textbf{of Design}} &\makecell[cc]{\textbf{Frequency Domain}\\\textbf{Input Constraint}}&\makecell[cc]{\textbf{Frame}\\\textbf{Length}\tnote{*}} &\makecell[cc]{\textbf{DoF}\\\textbf{Efficiency}}& \makecell[cc]{\textbf{References}}&\makecell[cc]{\textbf{Result}}\\ \hline

\multicolumn{2}{c|}{\makecell[cc]{\multirow{5}*{\rotatebox{90}{\textbf{Single-Component Schemes}$\mspace{20mu}$}}}}&\makecell[cc]{DCO-OFDM}  &\makecell[cc]{Adding DC bias\\(peak clipping is\\usually needed)}& \makecell[cc]{ Hermitian symmetry \\of $\textbf{X}$}&\makecell[cc]{$N$}&\makecell[cc]{$1$}&\makecell[cc]{[\ref{B94}, Sec. 5.3.2]\\\cite{CK96}}&\makecell[cc]{(\ref{DCO})}\\\cline{3-9}

\multicolumn{2}{c|}{~}&\makecell[cc]{ACO-OFDM} &\makecell[cc]{Clipping negative\\part to zero}&\makecell[cc]{Hermitian symmetry\\ of $\textbf{X}$, $X_k\equiv0$ for\\ even $k$}&\makecell[cc]{$N$}&\makecell[cc]{$\frac{1}{2}$}&\makecell[cc]{\cite{AL06}}&\makecell[cc]{(\ref{ACO}) \cite{Li1}}\\\cline{3-9}

\multicolumn{2}{c|}{~}&\makecell[cc]{PAM-DMT} &\makecell[cc]{Clipping negative\\part to zero}&\makecell[cc]{ Hermitian symmetry \\of $\textbf{X}$, $\textrm {Re}[X_k]\equiv0$}&\makecell[cc]{$N$}&\makecell[cc]{$\frac{1}{2}$}&\makecell[cc]{\cite{LRBK09}}&\makecell[cc]{\multirow{3}*{\makecell[cc]{\\ \\ \\(\ref{ACO})}}}\\\cline{3-8}

\multicolumn{2}{c|}{~}&\makecell[cc]{Flip-OFDM}&\makecell[cc]{Transmitting non-\\negative part and\\flipped negative part\\ separately}&\makecell[cc]{ Hermitian symmetry \\of $\textbf{X}$}&\makecell[cc]{$2N$}&\makecell[cc]{$\frac{1}{2}$}&\makecell[cc]{\cite{Yong07,FHV11,TSH12}}&{~}\\\cline{3-8}

\multicolumn{2}{c|}{~}&\makecell[cc]{Position modulating \\OFDM (PM-OFDM)}&\makecell[cc]{An analogy of\\Flip-OFDM} &\makecell[cc]{None}&\makecell[cc]{$4N$}&\makecell[cc]{$\frac{1}{2}$}&\makecell[cc]{\cite{NGHL12}}&{~}\\\hline

\makecell[cc]{\multirow{5}*{\rotatebox{90}{\textbf{Multiplexing based Schemes}$\mspace{65mu}$}}}&\makecell[cc]{\multirow{3}*{\rotatebox{90}{\textbf{Double-Component Schemes}$\mspace{-22mu}$}}}&\makecell[cc]{Asymmetrically\\clipped DC biased\\optical OFDM \\(ADO-OFDM) }&\makecell[cc]{Frequency division \\multiplexing (FDM)\\ of ACO- and DCO-\\OFDM}&\makecell[cc]{For DCO-OFDM, \\$X_k\equiv0$ for odd $k$} &\makecell[cc]{$N$}&\makecell[cc]{$1$}&\makecell[cc]{\cite{DPA11}}&\makecell[cc]{(\ref{ADO})}\\\cline{3-9}

~&~&\makecell[cc]{Hybrid asymmetri-\\cally clipped OFDM\\(HACO-OFDM) }&\makecell[cc]{FDM of ACO-\\OFDM and\\ PAM-DMT}&\makecell[cc]{For PAM-DMT, \\$X_k\equiv0$ for odd $k$} &\makecell[cc]{$N$}&\makecell[cc]{$\frac{3}{4}$}&\makecell[cc]{\cite{RK14}}&\multirow{2}*{\makecell[cc]{\\ \\(\ref{HACO}),\\Corollary 1}}\\\cline{3-8}

~&~&\makecell[cc]{Asymmetrically \\and symmetrically \\clipped OFDM\\ (ASCO-OFDM) }&\makecell[cc]{FDM of ACO- \\and Flip-OFDM\tnote{**}}&\makecell[cc]{For Flip OFDM, \\$X_k\equiv0$ for odd $k$} &\makecell[cc]{$2N$}&\makecell[cc]{$\frac{3}{4}$}&\makecell[cc]{\cite{WB15}}&{~}\\\cline{2-9}

~&{\multirow{2}*{\rotatebox{90}{\makecell[cc]{\textbf{Multi-Component}$\mspace{-12mu}$\\\textbf{Schemes}$\mspace{-12mu}$}}}}&\makecell[cc]{FDM-UOFDM\tnote{***}}&\makecell[cc]{FDM of $L$ \\ACO-OFDM \\components }&\makecell[cc]{Component $l$ uses \\the $2^{l-1}(2k+1)$-th\\subcarriers \\$\left(0\leq k \leq\frac{N}{2}-1\right)$}&\makecell[cc]{$N$}&$1-2^{-L}$&\makecell[cc]{\cite{EL14,LWEL,W15,ITH1}}&\makecell[cc]{(\ref{SEE}),\\Corollary 2}\\\cline{3-9}

~&~&\makecell[cc]{Enhanced unipolar \\OFDM (eU-OFDM)}&\makecell[cc]{Code division \\multiplexing of \\$L$ Flip-OFDM \\ components}&\makecell[cc]{No additional \\constraint}&\makecell[cc]{$2^LN$}&$1-2^{-L}$&\makecell[cc]{\cite{TVH15}}&\makecell[cc]{Theorem 6,\\ Corollary 2}\\\hline

\end{tabular}}
\begin{tablenotes}
        \footnotesize
        \item[*] Frame length stands for the length of a complete transmission period in terms of the number of channel uses in time domain.
        \item[**] Compared to \cite{WB15}, here we use an alternative, but equivalent description for the ASCO-OFDM.
        \item[***] It stands for several essentially equivalent schemes including the spectrally and energy efficient OFDM (SEE-OFDM)\cite{EL14},\\\cite{LWEL}, the layered ACO-OFDM (LACO-OFDM) \cite{W15}, and the eACO-OFDM \cite{ITH1}.
\end{tablenotes}
\end{threeparttable}
\end{table*}

\begin{table*}[tbp]
\renewcommand{\cellset}{\renewcommand{\arraystretch}{1.2}}
\centering
\caption{Unipolar OFDM Schemes for IM/DD}
\begin{threeparttable}
\scalebox{1}{
\begin{tabular}{l|l}
\hline

\makecell[cc]{\textbf{Basic Idea}\tnote{*}}&\makecell[cc]{\textbf{Examples}}\\\hline
\makecell[cl]{Resolving a block of original complex OFDM \\signal into four unipolar blocks} &\makecell[cc]{PM-OFDM \cite{NGHL12}}\\\hline
\makecell[cl]{Multiplexing (by superposition for example)\\two or more unipolar OFDM components} &\makecell[cc]{Schemes in \cite{DPA11,RK14,WB15,EL14,LWEL,W15,TVH15,ITH1}, ePAM-DMT\cite{ITH2}, EHACO-OFDM \cite{MChen},\\ RPO-OFDM\cite{EL13}, AHO-OFDM \cite{WWD15}, HOOK-ACO-OFDM \cite{YGL16}}\\\hline
\makecell[cl]{Using special discrete frequency domain input\\to generate unipolar OFDM signals} &\makecell[cc]{Spectral factorized optical OFDM \cite{AFH11}}\\\hline
\makecell[cl]{Using transforms other than discrete Fourier\\transform to generate unipolar OFDM signals} &\makecell[cc]{DHT based optical OFDM\cite{MMJ10},\\ FOFDM based on DCT \cite{Zhou2}}\\\hline
\makecell[cl]{Using specific transforms on time domain\\ OFDM signals to generate unipolar signals} &\makecell[cc]{POFDM\cite{EL141}, \\Asymmetrically reconstructed OOFDM \cite{SEUYou}}\\\hline
\makecell[cl]{Transmitting polarity information and absolute\\ values of original OFDM signals separately} &\makecell[cc]{PIC-flip-OFDM \cite{ZCZhang}, AAO-OFDM \cite{BWW17}}\\\hline
\makecell[cl]{Combining known unipolar schemes with\\ informative DC bias} &\makecell[cc]{DCIO-OFDM \cite{Xu}}\\\hline
\makecell[cl]{Combining known unipolar schemes with\\ nonlinear transforms} &\makecell[cc]{$\mu$-OFDM \cite{YZFG}}\\\hline
\makecell[cl]{Switching between known unipolar schemes} &\makecell[cc]{AAD-OFDM \cite{BUPT}}\\\hline

\end{tabular}}
\begin{tablenotes}
        \footnotesize
        \item[*] Note that DC bias is also used in some schemes.
\end{tablenotes}
\end{threeparttable}
\end{table*}

In recent years, numerous unipolar OFDM schemes have been proposed for IM/DD based optical communications, and Table II lists some representative examples.
As a direct and powerful way to improve the DoF efficiency, multiplexing two or more unipolar OFDM components has received particular attention.
Details of some multiplexing based unipolar OFDM schemes, which will be investigated in this paper,
are given in Table I. It is shown that the DoF efficiency loss of the unipolar design can be almost completely avoided by using frequency division multiplexing (FDM) or code division multiplexing (CDM) based multi-component schemes. Further studies on performance and optimization of unipolar OFDM can be found in extensive literatures, e.g., \cite{JS09,MEH11,DA13,TSH13,Lowery,SYG17}.
In particular, \cite{Lowery} and \cite{SYG17} compare performance of several multiplexing-based unipolar OFDM schemes.

Thus far, most studies on unipolar OFDM consider only the performance of \emph{uncoded} transmission, where the main performance metric is the bit error rate (BER) or the rate achieved for a target BER.
The current paper, alternatively, chooses \emph{information rate} as the main performance metric. 
For a given transmission scheme (specified by certain transceiver structures), its information rate is defined as the highest rate achieved by that scheme with arbitrarily low error probability as the channel coding length grows without bound. As a fundamental limit, the information rate depends neither on a particular target error probability (e.g., $10^{-3}$) nor on a particular channel code. Therefore, the information rate indicates the ultimate performance of unipolar OFDM schemes. More specifically, it approximates the maximum data rate achieved by a \emph{coded} unipolar OFDM system, with a sufficiently low error probability.

There are several existing information theoretic works on unipolar OFDM. The information rate of the ACO-OFDM was derived in \cite{Li1} and \cite{Li2} for average power constrained Gaussian optical intensity channels (OIC) with and without signal dispersion. In \cite{Yu1} and \cite{Yu2}, the information rates of the DCO-OFDM, the ACO-OFDM, and the Flip-OFDM were analyzed in light emitting diode (LED) based Gaussian OIC under average optical power and dynamic range constraints.
A framework for information rate maximization and parameter optimization (bias and clipping levels) of the ACO- and the DCO-OFDM in Gaussian OIC with nonlinear distortion was proposed in \cite{DH13}, with both electrical and optical power constraints (include average and dynamic range constraints) considered. Multiplexing based unipolar OFDM schemes, however, have not been considered in the aforementioned studies.

In \cite{Yu1,Yu2,DH13}, achievability results on an OFDM subcarrier were obtained by treating the distortion from clipping and other effects as independent additive noise, employing the capacity formula of the additive white Gaussian noise (AWGN) channel, and replacing the signal-to-noise-ratio (SNR) therein by the signal-to-noise-and-distortion-ratio (SNDR). Thus the obtained achievable rate for complex-valued input is as
\begin{equation}
\label{CAWGN}
R=\log(1+\textrm{SNDR}).
\end{equation}
From an information theoretic view, however, the achievability of (\ref{CAWGN}) needs further justification.
It is well known that the capacity formula of an AWGN channel gives the maximum achievable rate of average power constrained signaling corrupted by \emph{signal-independent} Gaussian noise, and Gaussian signaling achieves this maximum rate. Unfortunately, the distortion part in the evaluation of SNDR in (\ref{CAWGN}) is indeed a signal-dependent non-Gaussian noise. In fact, the rate given by (\ref{CAWGN}) is not equal to the maximum input-output mutual information of the considered frequency domain equivalent channel.
Hence, rigorous information theoretic results on unipolar OFDM with nonlinear distortion still needs to be established. In addition, the SNDR based results do not provide clues on how to design appropriate transceivers to achieve the rate as (\ref{CAWGN}).

Furthermore, although the capacity of OIC has been studied in several works \cite{HK04,LMW09,FH09,FH10,WHWCW,CMA15,ZZ17},
few studies have considered comparisons between information rates of unipolar OFDM schemes and the capacity of OIC.
The design of unipolar OFDM often leads to a significant loss of information rate. For example, the rate loss caused by a suboptimal DoF efficiency increases without a bound as SNR increases. In contrast, in electrical Gaussian channels both OFDM and serial transmission approach capacity \cite{FU98}.
Evaluating the rate losses of unipolar OFDM schemes for practical SNR values is thus necessary, especially for multiplexing based schemes since they
achieve higher DoF efficiencies at the price of increased complexity.
In \cite{DH13}, information rates of the DCO- and the ACO-OFDM were compared with the capacity of Gaussian OIC under dynamic range constraints. Our comparative study will consider more schemes including several multiplexing based ones.
For the capacity of discrete-time Gaussian OIC, although no analytic expression is known, tight bounds have been established for several common types of constraints \cite{LMW09,FH09,FH10,WHWCW,CMA15}. In addition, accurate evaluation can be numerically achieved using techniques pioneered in \cite{Smith71} when the input optical intensity is further bounded from above due to a peak power constraint. These results provide benchmarks for evaluating the performance of unipolar OFDM schemes.

\subsection{Summary of Contributions}

In this paper,
we study information rates of unipolar OFDM schemes listed in Table I in discrete-time Gaussian OIC under average optical power constraint. Table I includes the first five single-component schemes and several multiplexing based schemes, which include three double-component schemes and two types of multi-component schemes. Multiplexing based schemes in \cite{EL13,WWD15,YGL16} are not considered since they are more suitable for channels with dimming/lighting constraints.
The meaning of ``comparative study'' in this paper is two-fold: the study includes not only comparisons among information rates of unipolar schemes, but also comparisons between these information rates and the capacity of the considered channel.
Our main contributions are summarized as follows.
\begin{itemize}
\item
    For the DCO-OFDM and the ADO-OFDM, we derive information rate lower bounds which are achieved by employing independent and identically distributed (i.i.d.) complex Gaussian (ICG) or i.i.d. Gaussian (IG) codebook ensemble at the transmitter and nearest neighbour decoder
    at the receiver. Our derivations utilize information theoretic results on \emph{mismatched decoding}, especially the so-called \emph{generalized mutual information (GMI)} \cite{GLT,LS,Boss}.
    For the SNDR based information rate results as (\ref{CAWGN}), we show that our lower bounds coincide with them, thus establishing their achievability rigorously, and providing key engineering insights
    by connecting the information rates with specific transceiver design. The obtained lower bounds are tight in the sense of ``typical'' ICG/IG codebooks (see Sec. III for details).
    We also provide results on optimization of the DCO-OFDM and the ADO-OFDM for information rates maximization.
\item For the ACO-OFDM, the PAM-DMT, the Flip-OFDM, and the PM-OFDM,
    we derive their information rates under ICG/IG input in a unified way, and establish the equivalence of their information rates.
\item For several double- and multi-component schemes, we derive achievability results, including some asymptotic ones, based on successive decoding, interference cancellation, and results for single-component schemes. For the HACO-OFDM and the ASCO-OFDM, and for all multi-component schemes we considered, our results establish the equivalence of their information rates.
    We further give optimal power allocation parameters for double-component schemes numerically, which show that at low SNR the optimal strategy is allocating all power to the ACO-OFDM component (thus reducing into the single-component scheme).
\item Based on our results, we show that 1) at low SNR,
there is no need to use multiplexing based schemes,
but there are still considerable gaps between the information rates of single-component schemes and the channel capacity; 2) at high SNR, component-multiplexing provides great benefits, and the considered multi-component schemes approach the high-SNR capacity of the discrete-time Gaussian OIC under average optical power constraint to within 0.07 bits, by using a simple power allocation strategy we find.
\end{itemize}

\subsection{Organization and Notation}

The remaining part of this paper is organized as follows. Sec. II gives some preliminaries of our study. Sec. III, IV, and V study single-, double-, and multi-component schemes, respectively. Sec. VI presents numerical results and their discussions. Some concluding remarks are given in the last section.

Throughout the paper we use $N$ to denote the number of OFDM subcarriers.
We use $\textbf{F}$ and $\textbf{F}^{-1}$ to denote the discrete Fourier transform (DFT) matrix and the inverse discrete Fourier transform (IDFT) matrix, respectively, i.e.,
$\left[\textbf{F}\right]_{n,k}=N^{-\frac{1}{2}}\exp\left(-j2\pi k \frac{n}{N} \right)$,
$\left[\textbf{F}^{-1}\right]_{k,n}=N^{-\frac{1}{2}}$ $\exp\left(j2\pi n \frac{k}{N} \right)$,
where $0\leq k\leq N-1$ and $0\leq n\leq N-1$. Some upper case Roman letters including $X$, $S$, $C$, and $D$, and their bold forms, are used to denote scalar and vector signals in discrete frequency domain, respectively; the corresponding lower case Roman letters $x$, $s$, $c$, and $d$, and their bold forms, are used to denote scalar and vector signals in discrete time domain, respectively. Specifically, we use $s$ to denote the optical intensity. The two domains are connected by DFT/IDFT, e.g., $\textbf{x}=\textbf{F}^{-1}\textbf{X}$, where $[\textbf{x}]_n=x_n$ and $[\textbf{X}]_k=X_k$.  We use $\textrm E[\cdot]$ to denote expectation. We use ${\cal N}(a,b)$ and ${\cal {CN}}(a,b)$ to denote Gaussian and complex Gaussian distributions, respectively, with mean $a$ and variance $b$. The truncated Gaussian distribution with probability density function
$f_X(x)=\frac{1}{\sqrt{2\pi}\sigma}\exp\left( \frac{-x^2}{2\sigma^2}\right)$,
$x>0$, and $\textrm{Pr}(x=0)$ $=\frac{1}{2}$,
is denoted as ${\cal{TG}}\left(0,\sigma^2\right)$, and its mean is $\textrm E[x]=\frac{\sigma}{\sqrt{2\pi}}$ \cite{Li1}.
We use ${\cal C}_\textrm{AWGN}(\gamma)$ to denote the capacity of a complex-valued discrete-time AWGN channel $y=x+w$ with SNR $\gamma=\frac{\textrm E[x^2]}{\textrm E[w^2]}$, i.e.,
${\cal C}_\textrm{AWGN}(\gamma)=\log(1+\gamma)$.
The complex conjugate of a complex number $A$ is denoted as $\bar{A}$. For a matrix (or vector) $\textbf{A}$, $\textbf{A}^\textrm T$ stands for its transpose, $\textbf{A}^\textrm H$ stands for its conjugate transpose, and $\bar{\textbf{A}}$ stands for its element-by-element conjugation.
\section{Preliminaries}

The discrete-time Gaussian OIC considered in this paper is
\begin{equation}
\label{eqn:DTOIC}
r_n=s_n+z_n,\mspace{8mu}s_n\ge 0,
\end{equation}
where $z_n$ is i.i.d. Gaussian with variance $\sigma_z^2$. The transmitted optical intensity $s_n$ satisfies an average {optical} power constraint as
\begin{equation}
\label{AP}
\textrm E \left[s_n\right]\leq {\cal E}.
\end{equation}
Note that in (\ref{eqn:DTOIC}) we normalize the channel gain (including, e.g., the detector responsivity) to unity without loss of generality.
For brevity, in the remaining part of this paper, unless otherwise stated, ``Gaussian OIC'' always denotes its discrete-time version as (\ref{eqn:DTOIC}), ``frequency/time domain'' always refers to discrete frequency/time domain, and ``power'' always refers to optical power.

The above channel model has been widely used in information theoretic studies on OWC \cite{LMW09,FH09,FH10,WHWCW,CMA15}. In these studies, and also in the studies on unipolar OFDM, several more input constraints have been considered, including peak power constraint, dynamic range constraint \cite{Yu1,Yu2,DH13,DSH121}, dimming/lighting constraint (for visible light communications) \cite{EL13,WWD15,BUPT,WHWCW,Xu}, electrical power constraint \cite{DH13,DSH121}, and possible combinations of several single constraints.

In this paper, however, we only consider an average power constraint as (\ref{AP}), and pay more attention to the following unique problems in studying performance of unipolar OFDM:
\begin{itemize}
  \item Improving the DoF efficiency (also the high SNR performance) by component multiplexing.
  \item Evaluating the information rate by studying the frequency domain equivalent channel and possible interferences between subchannels.
  \item Further boosting the information rate by parameter optimization.
\end{itemize}
In contrast, achieving the capacity of Gaussian OIC only requires optimizing the distribution of $s$ with given constraints. So for unipolar OFDM,
we believe that considering the above problems should precede considering more constraints.
Note that if we alternatively assume a peak power constraint as $\textrm{Pr}(s_n>{\cal A})=0$, then bipolar OFDM schemes can be employed by adding a DC bias ${\cal A}/2$ without cost, and therefore we do not have to study unipolar design.
By considering the average power constraint, we can identify the impact of unipolar design on performance from that of various further constraints in practical systems. In fact, in this case closed-form information rate results can be obtained for most unipolar OFDM schemes we considered. The extension of our study to OIC with more constraints will be briefly discussed in Sec. VII.

Throughout the paper we make the following assumptions on the transceiver of unipolar OFDM, and denote the information rate derived under these assumptions by ${\cal R}_\textrm{OFDM}$.
\begin{itemize}
  \item The receiver uses decoding techniques as that commonly used in practical OFDM based IM/DD systems (not necessarily optimal). Furthermore, we assume that a nearest neighbor decoder is used.
  \item The codebook of the frequency domain input is generated according to an ICG/IG ensemble.
\end{itemize}

We make the first assumption for achieving a valid performance evaluation of practical unipolar OFDM systems. For example, the receiver of the ACO-OFDM usually discards the received signal on even subcarriers, and performs minimum Euclidean distance decoding for odd subcarriers. Therefore, the input-output mutual information ${\cal I}_\textrm{OFDM}$, which may be higher than ${\cal R}_\textrm{OFDM}$, is not achievable since practical decoders are not optimal.
Calculating the single-letter input-output mutual information ${\cal I}(s;r)$ under the distribution of the transmitted optical intensity $s$ (e.g., the study on the ASCO- and the ADO-OFDM in \cite{Wu}) is also inappropriate here
because $s$ is not i.i.d. and therefore the single-letter characterization of information rate is not valid from an information theoretic perspective. In fact, when the DoF efficiency of a unipolar OFDM scheme is smaller than one, the pre-log factor of ${\cal R}_\textrm{OFDM}$ should be smaller than $\frac{1}{2}$, while ${\cal I}(s;r)$ has the form $\frac{1}{2}\log(1+a\frac{{\cal E} ^2}{\sigma_z^2})$.

We make the second assumption for several reasons:
1) It enables the use of the GMI framework in its simplest form \cite{Boss} for the DCO- and the ADO-OFDM, and it leads to concise close-form results for other schemes.
2) In Gaussian channels the performance loss due to using practical PAM/QAM constellations other than Gaussian/complex Gaussian input is limited \cite{FU98}.

To obtain neat expressions, all our results are asymptotic results for large $N$. Note that for practical values of $N$ (e.g., $N\ge 64$) these results maintain high accuracy, and exact results can be easily obtained from our derivation. For brevity, detailed descriptions of the considered unipolar OFDM schemes are combined into the proof of our results.

\begin{figure*}[b]
\begin{align}
\label{DCO}
{\cal R}_\textrm{DCO-OFDM}=
\max_{\nu>0}\frac{1}{2}\log\bigg(1+\frac{\textrm{erf}^2\left(\nu{\cal E}\right)}{\textrm{erf}\left(\nu{\cal E}\right)- \textrm{erf}^2\left(\nu{\cal E}\right)-2\pi^{-\frac{1}{2}}\nu{\cal E}\exp\left(-\nu^2{\cal E}^2\right)+2\nu^2{\cal E}^2\textrm{erfc}\left(\nu{\cal E}\right)+2\nu^2{\sigma_z^2}}\bigg).
\tag{4}
\end{align}
\end{figure*}
\section{Single-Component Schemes}

This section considers the first five unipolar OFDM schemes in Table I.
In general, we derive information rates of those unipolar schemes as follows: 1) find the equivalent channel model of a unipolar OFDM scheme in frequency domain as $\{Y_k=X_k+Z_k\}$, $k\in \cal K$, ($\cal K$ denotes the set of subcarriers carrying independent input symbols, e.g., for the DCO-OFDM ${\cal K}=\{1,...,N/2-1\}$) and derive its information rate with respect to $\sigma_X^2$; 2) determine the relationship between $\sigma_X^2$ and the average optical power constraint $\cal E$ (called the $\sigma_X$-$\cal E$ relationship), according to the distribution of the optical intensity transmitted.

We utilize the general theoretical framework for transmission with transceiver distortion proposed in \cite{Boss} to study the information rate of the DCO-OFDM. This framework is based on the GMI (denoted as ${\cal I}_\textrm{GMI}$) which is a lower bound on the information rate of a communication scheme with mismatched decoding (i.e., the receiver using a given decoding metric with is suboptimal) \cite{GLT,LS}. This GMI-based framework is briefly explained in Appendix A, in which we also extend its original version to vector channels. Moreover, the GMI is the \emph{highest} information rate below which the average probability of error with that decoding metric, further averaged over the chosen ensemble of codebooks, converges to zero as the code length tends to infinity. Therefore our lower bound on the information rate of the DCO-OFDM is tight in the sense of ``typical'' ICG/IG codebooks.

\emph{Theorem 1}: For the DCO-OFDM with ICG codebook ensemble and nearest neighbor decoding, an achievable information rate ${\cal R}_\textrm{DCO-OFDM}$ is given by (\ref{DCO}).

\begin{IEEEproof}Consider a block of the input of the DCO-OFDM as
\setcounter{equation}{4}
\begin{equation}
\textbf{X}=\left[0,X_1,...,X_{\frac{N}{2}-1},0,\bar{X}_{\frac{N}{2}-1},...,\bar{X}_1\right]^\textrm T
\end{equation}
which is a length-$N$ complex vector with Hermitian symmetry. For $1\leq k \leq \frac{N}{2}-1$, let $X_k$ be i.i.d. and $X_k\sim {\cal {CN}}\left(0,\sigma_X^{2}\right)$. Taking IDFT of $\textbf{X}$,
we obtain
$\textbf{x}=\left[x_1,...,x_N\right]^\textrm T
=\textbf{F}^{-1}\textbf{X}$.
Since IDFT is unitary, we have $\|\textbf{x}\|=\|\textbf{X}\|$. Moreover, it can be shown that
\begin{equation}
\label{Exx}
\textrm E\left[x_n x_{n^\prime}\right]=
\begin{cases}
\frac{N-2}{2N}\sigma_X^2,&n={n^\prime},\\
0,&n-{n^\prime}\mspace{8mu}\textrm{is odd,}\\
-\frac{1}{N}\sigma_X^2,&n-{n^\prime}\mspace{8mu}\textrm{is even}, \mspace{8mu}n-{n^\prime}\neq0.
\end{cases}
\end{equation}
So $x_n$ satisfies $x_n\sim {\cal N}\left(0,\sigma_x^{2}\right)$ for given $n$ since it is a linear combination of real and imaginary parts of $X_k$ (both are i.i.d. Gaussian variables), where
$\sigma_x^{2}=\frac{N-2}{N}\sigma_X^{2}$,
and $x_n$ is asymptotically i.i.d. as $N\rightarrow\infty$.
The unipolar input to the OIC, $\textbf{s}=[s_1,...,s_{N-1}]^\textrm T$, is obtained by clipping the signal peaks symmetrically as
\begin{equation}
\label{clip}
\begin{split}
c_n=
\begin{cases}
{\cal A} \mspace{8mu} &x_n> {\cal A}\\
x_n,  &|x_n|\leq {\cal A}\\
-{\cal A},  &x_n< -{\cal A}
\end{cases}
\end{split}
\end{equation}
and adding a DC bias
${\cal A}$ on $c_n$. Apparently $\textrm E\left[s_n\right]={\cal A}$, and we let ${\cal A} ={\cal E}$.

The output of the OIC after removing the DC bias is
$\textbf{y}=\textbf{c}+\textbf{z}$
where $[\textbf{y}]_n=y_n$.
By taking DFT of $\textbf{y}$ we obtain
$\textbf{Y}=\textbf{C}+\textbf{Z}=\textbf{F}\textbf{c}+\textbf{Z}$
where $[\textbf{Y}]_k=Y_k$, $1\leq k\leq N$.
Note that $\textbf{C}$ is determined by $\textbf{X}$ via (\ref{clip}). Without loss of optimality, we consider the $(\frac{N}{2}-1)$ dimensional equivalent channel
\begin{equation}
\label{v1YCZ}
\textbf{\textsf{Y}}=\textbf{\textsf{C}}+\textbf{\textsf{Z}}
\end{equation}
where $\textbf{\textsf{Y}}=[Y_1,...,Y_{\frac{N}{2}-1}]^\textrm T$, $\textbf{\textsf{C}}$ is determined by $\textbf{\textsf{X}}=[X_1,...,X_{\frac{N}{2}-1}]^\textrm T$ and consists of asymptotically i.i.d. elements, and the noise $\textbf{\textsf{Z}}$ consists of i.i.d. elements.

For transmission at rate $\cal R$, assume that a message $m$ is selected from ${\cal M} =\left\{1,...,\lfloor \exp(\ell \cal R)\rfloor\right\}$ uniformly randomly. The encoder maps $m$ to a length-$\ell$ codeword $\left[\textbf{\texttt{X}}^{(1)}(m),...,\textbf{\texttt{X}}^{(\ell)}(m)\right]^\textrm T$
in an ICG codebook ensemble, where $\textbf{\texttt{X}}^{(i)}(m)=\left[\texttt{X}^{(i)}_1(m),...,\texttt{X}^{(i)}_{N/2-1}(m)\right]^\textrm T$ for $1\leq i\leq \ell$. At the receiver, we let the decoder follow a (scaled) nearest neighbor decoding rule as
\begin{equation}
\label{D}
\hat{m}=\mathop{\arg\min}_{\textsf{m}\in{\cal M}}\frac{1}{\ell}\sum_{i=1}^{\ell}\left\|\textbf{\textsf{Y}}^{(i)}-a\textbf{\texttt{X}}^{(i)}(\textsf{m})\right\|^2,\mspace{8mu} \textsf{m}\in {\cal M},
\end{equation}
where $\|\cdot\|$ is the ${\cal L}_2$ norm.
In (\ref{D}), $a$ is a decoding scaling parameter to be optimized.
Under the above assumption, the GMI achieved by the channel $\textbf{\textsf{Y}}=\textbf{\textsf{C}}+\textbf{\textsf{Z}}$ in (\ref{v1YCZ}) is (see Appendix A)
\begin{equation}
{\cal I}_\textrm{GMI}=\left(\frac{N}{2}-1\right)\log\left(1+\frac{\Delta}{1-\Delta}\right),
\end{equation}
where
\begin{equation}
\label{Delta}
\Delta=\frac{\left|\textrm E\left[\textbf{\textsf{Y}}^\textrm H \textbf{\textsf{X}}\right]\right|^2}{\left(N/2-1\right)\sigma_X^2\left(\textrm E\left[\|\textbf{\textsf{C}}\|^2\right]+\textrm E\left[\|\textbf{\textsf{Z}}\|^2\right]\right)},
\end{equation}
and the expectation $\textrm E\left[\cdot\right]$ is taken with respect to $\textbf{\textsf{X}}$ and $\textbf{\textsf{Z}}$. Here the decoding scaling parameter $a$ is optimally set as
\begin{equation}
\label{aopt}
a_\textrm{opt}=\frac{\textrm E\left[\textbf{\textsf{Y}}^\textrm H \bar{\textbf{\textsf{X}}}\right]}{\left(N/2-1\right)\sigma_X^2}.
\end{equation}

To evaluate $\Delta$ and $a_{\textrm{opt}}$, we write $c_n$ as
$c_n=\alpha x_n+d_n$ where
$\alpha=\frac{\textrm E\left[c_n x_n\right]}{\sigma_x^2}$ and
$d_n$ satisfies $\textrm E\left[d_n x_n\right]=0$.
By taking DFT
we obtain
$C_k=\alpha X_k+D_k$,
where
$D_k=\frac{1}{\sqrt N} \sum_{n=1}^N d_n\exp\left(-j2\pi n\frac{k}{N} \right)$
which satisfies
$\textrm E \left[|D_k|^2\right]=\textrm E \left[d_n^2\right]$.
Moreover, we have
$\textrm E\left[\bar{D}_k X_k\right]=0$,
which is because
\begin{equation}
\textrm E \left[\textbf{XD}^\textrm H\right]=\textrm E \left[\textbf{F}\textbf{x}(\textbf{F}\textbf{d})^\textrm H\right]=\textrm E \left[\textbf{F}\textbf{x}\textbf{d}^\textrm H\textbf{F}^{-1}\right]=\textbf{0}
\end{equation}
since $\textrm E\left[x_n d_{n^\prime}\right]=0$ for any $n,n^\prime$.
Substituting (\ref{v1YCZ}) into (\ref{aopt}) and (\ref{Delta}) yields
\begin{align}
a_\textrm{opt}&=\frac{\textrm E\left[(\textbf{\textsf{C}}+\textbf{\textsf{Z}})^\textrm H \textbf{\textsf{X}}\right]}{\left(N/2-1\right)\sigma_X^2}\notag\\
&=\frac{\sum_{k=1}^{N/2-1}\textrm E\left[\alpha \bar{X}_k X_k+\bar{D}_k X_k+\bar{Z}_kX_k\right]}{\left(N/2-1\right)\sigma_X^2}\notag\\
&=\frac{\alpha\sum_{k=1}^{N/2-1}\textrm E\left[ \bar{X}_k X_k\right]}{\left(N/2-1\right)\sigma_X^2}\notag\\
&=\alpha,
\end{align}
and
\begin{align}
\Delta &=\frac{\left|\textrm E\left[(\textbf{\textsf{C}}+\textbf{\textsf{Z}})^\textrm H \textbf{\textsf{X}}\right]\right|^2}{\left(N/2-1\right)\sigma_X^2\left(\textrm E\left[\|\textbf{\textsf{C}}\|^2\right]+\textrm E\left[\|\textbf{\textsf{Z}}\|^2\right]\right)}\notag\\
&=\frac{\left(\left(N/2-1\right)\alpha\sigma_X^2\right)^2}{\left(N/2-1\right)\sigma_X^2\sum_{n=1}^{N/2-1}\left(\textrm E\left[|C_n|^2\right]+\textrm E\left[|Z_n|^2\right]\right)}\notag\\
&=\frac{\left(N/2-1\right)\alpha^2\sigma_X^2}{\sum_{n=1}^{N/2-1}\left(\alpha^2\textrm E\left[|X_n|^2\right]+\textrm E\left[|D_n|^2\right]+\textrm E\left[|Z_n|^2\right]\right)}\notag\\
&=\frac{\alpha^2\sigma_X^2}{\alpha^2\sigma_X^2+\textrm E\left[|D_n|^2\right]+\sigma_z^2}.
\end{align}
We thus obtain
\begin{equation}
\label{SNDR}
\frac{\Delta}{1-\Delta}=\frac{\alpha^2\sigma_X^2}{\textrm E\left[d_n^2\right]+\sigma_z^2}.
\end{equation}
Note that ${\cal I}_\textrm{GMI}$ measures the information rate per \emph{frequency domain} channel use, and in time domain the channel is used for $N$ times. Then we obtain an achievable information rate as
\begin{equation}
\label{GMIbound}
{\cal R}_\textrm{DCO-OFDM}=\frac{1}{N}{\cal I}_\textrm{GMI}=\frac{N-2}{2N}\log\left(1+\frac{\alpha^2\sigma_X^2}{\textrm E\left[d_n^2\right]+\sigma_z^2}\right).
\end{equation}
According to (\ref{clip}), by simple calculation we obtain
$\textrm E\left[c_n x_n\right]=\textrm {erf}\left(\frac{\cal E}{\sqrt {2}\sigma _x}\right)\sigma_x^2$
which implies
\begin{equation}
\label{alpha}
\alpha=\textrm {erf}\left(\frac{\cal E}{\sqrt {2}\sigma _x}\right),
\end{equation}
and the variance of $c_n$ as
\begin{align}
\label{vcn}
\textrm E\left[c_n^2\right]=\sigma_x^2\left( \textrm{erf}\left(\frac{{\cal E}}{\sqrt {2}\sigma_x}\right)-\sqrt{\frac{2}{\pi}}\frac{{\cal E}}{\sigma _x}\exp\left(-\frac{{\cal E}^2}{2\sigma _x^2}\right)\right)\notag\\
+{\cal E}^2\textrm{erfc}\left(\frac{{\cal E}}{\sqrt {2}\sigma_x}\right).
\end{align}
We can thus obtain $\textrm E\left[d_n^2\right]$ by noting that
\begin{equation}
\label{vdn}
\textrm E \left[c_n^2\right]=\textrm E \left[\alpha^2 x_n^2+d_n^2+2\alpha x_n d_n\right]\\
=\alpha^2\sigma_x^2+\textrm E\left [d_n^2\right].
\end{equation}
Combining (\ref{GMIbound})--(\ref{vdn}), letting $\nu=\frac{1}{\sqrt2\sigma_X}$ noting that $\sigma_x^{2}=\frac{N-2}{N}\sigma_X^{2}$, letting $N\rightarrow\infty$, and choosing $\nu$ to maximize the RHS of (\ref{GMIbound}), we obtain (\ref{DCO}).
\end{IEEEproof}

\emph{Note}: In the above derivation a decomposition $c_n=\frac{\textrm E\left[c_n x_n\right]}{\sigma_x^2} x_n+d_n$, which is a discrete-time analogy of the Bussgang decomposition \cite{Bussgang}, is useful. This decomposition is exactly the key step in the SNDR based rate analysis. For example, see \cite{OI} for clipped OFDM in (electrical) Gaussian channels, and \cite{Yu1,Yu2,DH13} for unipolar OFDM in Gaussian OIC under other types of input power constraints.
In fact, our result will coincide with the SNDR based result if the same input power constraint is used. Thus, our GMI based information rate lower bound, which is obtained from the random coding analysis on the decoding error probability, validates the SNDR based rate analysis, and further establishes the connection between the information rate result and specific transceiver design.\footnote{The achievability of the SNDR based result can also be proved using the fact that the Guassian noise is the worst \emph{uncorrelated} noise, which is given in \cite{HH}.}

\begin{figure*}[b]
\begin{align}
\label{ADO}
&{\cal R}_\textrm{ADO-OFDM}=\max_{0\leq\lambda\leq1,\nu>0}\frac{1}{4}\log\left(\left(1+\frac{\pi(1-\lambda)^2{\cal E}^2}{ (2\nu^2)^{-1}{\cal D}+\sigma_z^2}\right)\cdot
\left(1+\frac{2\textrm{erf}^2\left({\nu\lambda{\cal E}}\right)}{ {\cal D}+2\nu^2\sigma_z^2}\right)\right),\notag\\
\textrm{where}\mspace{8mu}
&{\cal D}=\textrm{erf}\left(\nu\lambda{\cal E}\right)- \textrm{erf}^2\left(\nu\lambda{\cal E}\right)-2\pi^{-1/2}\nu\lambda{\cal E}\exp\left(-\nu^2\lambda^2{\cal E}^2\right)+2\nu^2\lambda^2{\cal E}^2\textrm{erfc}\left(\nu\lambda{\cal E}\right).
\tag{22}
\end{align}
\end{figure*}
\begin{figure*}[t]
\begin{align}
\label{d6}
\textrm E \left[d_{2,n}^2\right]=
\sigma_{x2}^2\left( \textrm{erf}\left(\frac{\lambda{\cal E}}{\sqrt {2}\sigma_{x2}}\right)- \textrm{erf}^2\left(\frac{\lambda{\cal E}}{\sqrt {2}\sigma_{x2}}\right)-\sqrt{\frac{2}{\pi}}\frac{\lambda{\cal E}}{\sigma _{x2}}\exp\left({-\frac{\lambda^2{\cal E}^2}{2\sigma _{x2}^2}}\right)\right)
+\lambda^2{\cal E}^2\textrm{erfc}\left(\frac{\lambda{\cal E}}{\sqrt {2}\sigma_{x2}}\right)
\tag{26}
\end{align}
\end{figure*}

\emph{Theorem 2}: The information rates of the ACO-OFDM, the Flip-OFDM, and the PM-OFDM, with ICG codebook ensemble and nearest neighbor decoding, are all given by
\begin{equation}
\label{ACO}
{\cal R}=\frac{1}{4}\log\left(1+\frac{\pi{\cal E}^2}{\sigma_z^2}\right),
\end{equation}
and the information rate of the PAM-DMT with IG codebook ensemble and nearest neighbor decoding is also given by (\ref{ACO}).

\emph{Note}: The fact that the ACO-OFDM with complex Gaussian input achieves the information rate as (\ref{ACO}) has been pointed out in \cite{Li1}, while the fact that these four schemes perform equivalently in terms of other performance metrics (e.g., BER) has also been observed, see \cite{TSH13}.

\begin{IEEEproof} We give a unified proof for all four schemes, and details of each scheme are given in Table III.

Consider a length-$N$ block of the input symbols for each unipolar OFDM scheme, denoted as $\textbf{X}$. We always let $X_k$ be i.i.d. (with the exception of the Hermitian symmetry).
Taking IDFT of $\textbf{X}$ yields $\textbf{x}$ satisfying $x_n\sim{\cal N}\left(0,\sigma_x^2\right)$. A frame of the unipolar input to the OIC is obtained by a transform as $\textbf{s}={\textsf T}(\textbf{x})$.
For the first three schemes the transmit optical intensity satisfies $s\sim$ ${\cal{TG}}\left(0,\sigma_x^2\right)$ and $\textrm E\left[s\right]={\cal E}=\frac{\sigma_x}{\sqrt{2\pi}}$, and for the PM-OFDM it satisfies $s\sim{\cal{TG}}\left(0,\frac{1}{2}\sigma_x^2\right)$ and $\textrm E[s]=$ ${\cal E}=\frac{\sigma_x}{2\sqrt{\pi}}$.
Noting that $\|\textbf{x}\|=\|\textbf{X}\|$, the $\sigma_X$-$\cal E$ relationship can be obtained based on $\textsf T$.

\begin{table*}[tbp]
\renewcommand{\cellset}{\renewcommand{\arraystretch}{1.5}}
\centering
\caption{Details in the Proof of Theorem 2}
\begin{threeparttable}
\scalebox{0.92}{
\begin{tabular}{l|l|l|l|l|l|l}
\hline
\makecell[cc]{\textbf{Name}}&\makecell[cc]{$\textbf{X}$}
&\makecell[cc]{$\textbf{s}=\textsf T(\textbf{x})$}&\makecell[cc]{$\sigma_X^2=$} & \makecell[cc]{$\textbf{y}=\textsf{R}(\textbf{r})$}&\makecell[cc]{\textbf{Frequency Domain}\\\textbf{Equivalent Channel}}&\makecell[cc]{ ${\textrm{SNR}_\textrm e}$}\\ \hline

\makecell[cl]{ACO-OFDM}  &\makecell[cl]{$\left[0,X_1,...,X_{\frac{N}{2}-1},0,\bar{X}_{\frac{N}{2}-1},...,\bar{X}_1\right]^\textrm T$\\for odd $k$, $X_k\sim {\cal{CN}}\left(0,\sigma_X^{2}\right)$\\for even $k$, $X_k\equiv0$}
&\makecell[cc]{$\textbf{s}=[s_1,...,s_{N-1}]^\textrm T$\\$s_n=\max(x_n,0)$}&\makecell[cc]{$\frac{2N}{N-2}{2\pi}{\cal E}^2$}&\makecell[cc]{$y_n=r_{n}$}&\makecell[cc]{$Y_k=\frac{1}{2}X_k+W_k$\\$1\leq k\leq \frac{N}{2}-1$,\\$k$ is odd}&\makecell[cc]{$\frac{\sigma_X^2}{4\sigma_z^2}$}\\\hline

\makecell[cl]{PAM-DMT}  &\makecell[cl]{$\left[0,X_1,...,X_{\frac{N}{2}-1},0,\bar{X}_{\frac{N}{2}-1},...,\bar{X}_1\right]^\textrm T$\\$\textrm {Re}[X_k]\equiv0$, $\textrm {Im}[X_k]\sim {\cal N}\left(0,\sigma_X^{2}\right)$}
&\makecell[cc]{$\textbf{s}=[s_1,...,s_{N-1}]^\textrm T$\\$s_n=\max(x_n,0)$}&\makecell[cc]{$\frac{N}{N-2}{2\pi}{\cal E}^2$}&\makecell[cc]{$y_n=r_{n}$}&\makecell[cc]{$\textrm{Im}[Y_k]=\frac{1}{2} \textrm{Im}[X_k]+$\\$\textrm{Im}[Z_k]$,$\mspace{4mu}$$1\leq k\leq \frac{N}{2}-1$}&\makecell[cc]{$\frac{\sigma_X^2}{2\sigma_z^2}$}\\\hline

\makecell[cl]{Flip-OFDM}  &\makecell[cl]{$\left[0,X_1,...,X_{\frac{N}{2}-1},0,\bar{X}_{\frac{N}{2}-1},...,\bar{X}_1\right]^\textrm T$\\$X_k\sim {\cal{CN}}\left(0,\sigma_X^{2}\right)$}
&\makecell[cc]{$\textbf{s}=\left[\textbf{s}_\textrm P^\textrm T,\textbf{s}_\textrm N^\textrm T\right]^\textrm T$,\\$s_{\textrm P,n}=\max(x_n,0)$,\\ $s_{\textrm N,n}=-\min(x_n,0)$}&\makecell[cc]{$\frac{N}{N-2}{2\pi}{\cal E}^2$}&\makecell[cc]{$y_n=r_{\textrm P,n}-r_{\textrm N,n}$\\
$=x_n+w_n$,\\$w_n\sim{\cal N}\left(0,2\sigma_z^2\right)$}&\makecell[cc]{$Y_k=X_k+W_k$\\$1\leq k\leq \frac{N}{2}-1$}&\makecell[cc]{$\frac{\sigma_X^2}{2\sigma_z^2}$}\\\hline

\makecell[cl]{PM-OFDM}  &\makecell[cl]{$\left[X_0,...,X_{N-1}\right]^\textrm T$, $X_k\sim {\cal{CN}}\left(0,\sigma_X^{2}\right)$}
&\makecell[cc]{$\textbf{s}=\left[\textbf{s}_1^\textrm T,\textbf{s}_2^\textrm T,\textbf{s}_3^\textrm T,\textbf{s}_4^\textrm T\right]^\textrm T$\\$s_{1,n}=\max\left(\textrm{Re}[x_n],0\right)$\\$s_{2,n}=-\min\left(\textrm{Re}[x_n],0\right)$\\$s_{3,n}=\max\left(\textrm{Im}[x_n],0\right)$
\\$s_{4,n}=-\min\left(\textrm{Im}[x_n],0\right)$}&\makecell[cc]{${4\pi}{\cal E}^2$}&\makecell[cc]{$y_n=r_{1,n}-r_{2,n}$\\$+j(r_{3,n}-r_{4,n})$\\$=x_n+w_n$,\\$w_n\sim{\cal CN}\left(0,4\sigma_z^2\right)$}&\makecell[cc]{$Y_k=X_k+W_k$\\$0\leq k\leq N$}&\makecell[cc]{$\frac{\sigma_X^2}{4\sigma_z^2}$}\\\hline

\end{tabular}}
\end{threeparttable}
\end{table*}

At the receiver, we obtain $\textbf{r}=\textbf{s}+\textbf{z}$ and transform it to a length-$N$ block as
$\textbf{y}=\textsf{R}(\textbf{r})$. Taking DFT, we obtain $\textbf{Y}=\textbf{Fy}$. We then discard certain elements of $\textbf{Y}$, including 1) conjugate elements (without loss of optimality), 2) for the ACO-OFDM those corresponding to even subcarriers for distortion noise cancelling, and 3) for the PAM-DMT the real part of $Y_k$. A set of parallel equivalent channels in frequency domain is thus obtained for each scheme (for the ACO-OFDM or the PAM-DMT, there is a scaling on the amplitude of $X_k$ due to the asymmetric clipping). The equivalent SNRs of these channels (denoted as $\textrm{SNR}_\textrm e$) can be obtained based on the distributions of channel input and noise. Since each equivalent channel is an AWGN channel, the information rate achieved by nearest neighbor decoding is $\frac{1}{2}{\cal C}_\textrm{AWGN}\left(\textrm{SNR}_\textrm e\right)$ for the PAM-DMT using IG codebook ensemble, and is ${\cal C}_\textrm{AWGN}\left(\textrm{SNR}_\textrm e\right)$ for the other three schemes using ICG codebook ensemble. For each scheme, (\ref{ACO}) can be obtained by scaling the corresponding capacity of an equivalent channel with the number of equivalent channels, the frame length, and the $\sigma_X$-$\cal E$ relationships, and letting $N\rightarrow\infty$.
\end{IEEEproof}

\section{Double-Component Schemes}

This section studies the ADO-OFDM, the HACO-OFDM, and the ASCO-OFDM. For all these cases, results are derived by decoding the ACO-OFDM component first, performing interference cancellation, and then decoding the second component. In particular, for the ADO-OFDM, we use a result from \cite{Lapidoth} to derive the information rate of the ACO-OFDM component which is corrupted by the clipping noise from the DCO-OFDM component, and use the GMI framework to derive the information rate of the DCO-OFDM component.

\emph{Theorem 3}: For the ADO-OFDM with ICG codebook ensemble and nearest neighbor decoding for both components, an achievable information rate ${\cal R}_\textrm{DCO-OFDM}$ is given by (\ref{ADO}).

\begin{IEEEproof}
Consider a block of the input of the ACO-OFDM component as
\setcounter{equation}{22}
\begin{align}
\textbf{X}_1=\big[&0,X_{1,1},0,X_{1,3}, ...,0,X_{1,\frac{N}{2}-1},\notag\\
&0,\bar{X}_{1,\frac{N}{2}-1},0,...,\bar{X}_{1,3},0,\bar{X}_{1,1}\big]^\textrm T
\end{align}
where $N$ is divisible by 4. For odd $k$ satisfying $1\leq k \leq \frac{N}{2}-1$, we let $X_{1,k}$ be i.i.d. and $X_{1,k}\sim $ $ {\cal{CN}}\left(0,\sigma_{X1}^{2}\right)$. Taking IDFT of $\textbf{X}_1$ yields $\textbf{x}_1$ satisfying $x_{1,n}\sim {\cal N}\left(0,\sigma_{x1}^{2}\right)$ for given $n$, where
$\sigma_{x1}^{2}=\frac{\sigma_{X1}^{2}}{2}$, and $x_{1,n}$ is asymptotically i.i.d. as $N\to \infty$.
The ACO-OFDM component, denoted as $\textbf{s}_1=[s_{1,1},...,s_{1,N-1}]^\textrm T$, is obtained by asymmetric clipping.
So $s_{1,n}\sim{\cal{TG}}\left(0,\sigma_{x1}^2\right)$
and the power cost is
$\textrm E\left[s_{1,n}\right]=\frac{\sigma_{x1}}{\sqrt{2\pi}}$.
Letting $\textrm E\left[s_{1,n}\right] =(1-\lambda){\cal E}$ we obtain the $\sigma_X$-$\cal E$ relationship of the ACO-OFDM component as
$\sigma_{X1}^2=(1-\lambda)^2{4\pi}{\cal E}^2$.

Consider a block of the input of the DCO-OFDM component as
\begin{align}
\textbf{X}_2=\big[&0,0,X_{2,2},0,X_{2,4},..., X_{2,\frac{N}{2}-2}, 0,\notag\\
&0,0, \bar{X}_{2, \frac{N}{2}-2}, ...,\bar{X}_{2,4},0,\bar{X}_{2,2},0\big]^\textrm T
\end{align}
For even $k$ satisfying $2\leq k \leq \frac{N}{2}-2$, let $X_{2,k}$ be i.i.d. and $X_{2,k}\sim {\cal{CN}}\left(0,\sigma_{X2}^{2}\right)$. Taking IDFT of $\textbf{X}_2$ yields $\textbf{x}_2$ satisfying $x_{2,n}\sim {\cal N}\left(0,\sigma_{x2}^{2}\right)$ for given $n$, where
$\sigma_{x2}^{2}=\frac{N-4}{2N}\sigma_{X2}^{2}$, and $x_{2,n}$ is asymptotically i.i.d. as $N\to \infty$.
The DCO-OFDM component, denoted as $\textbf{s}_2=[s_{2,1},...,s_{2,N-1}]^\textrm T$, is obtained by clipping the peaks of $\textbf{x}_2$ symmetrically as in (\ref{clip}) (in the sequel, the obtained signal is denoted as $\textbf{c}_2$), and adding a DC bias, i.e., $\textbf{s}_2=\textbf{c}_2+\cal A$. Apparently
$\textrm E\left[s_{2,n}\right]={\cal A}$,
and we let ${\cal A} =\lambda{\cal E}$.
A frame of the unipolar ADO-OFDM input to the OIC is thus
$\textbf{s}=\textbf{s}_1+\textbf{s}_2$
where
$\textrm E\left[s_{n}\right]={\cal E}$.

At the receiver, removing the DC bias from the received signal and then taking its DFT, we obtain $\frac{N}{2}-1$ parallel channels as
\begin{equation}
\label{YHXZ6-7}
\begin{split}
Y_{k}=
\begin{cases}
\frac{1}{2} X_{1,k}+Z_k+D_{2,k}, \mspace{8mu} &k \mspace{8mu}\textrm {is}\mspace{8mu}\textrm {odd},\\
C_{2,k}+Z_k+D_{1,k}, \mspace{8mu} &k \mspace{8mu}\textrm {is}\mspace{8mu}\textrm {even},
\end{cases}
\end{split}
\end{equation}
where $1\leq k \leq \frac{N}{2}-1$, $D_{1,k}$ and $D_{2,k}$ are the distortion terms introduced by the ACO-OFDM component and the DCO-OFDM component, respectively.
The distortion $D_{2,k}$
satisfies
$\textrm E \left[|D_{2,k}|^2\right]=\textrm E \left[d_{2,n}^2\right]$,
where $\textrm E \left[d_{2,n}^2\right]$ is given in (\ref{d6}).
Note that there are $\frac{N}{4}$ odd numbered channels and $\frac{N}{4}-1$ even numbered channels in (\ref{YHXZ6-7}).

We first consider the information rate of the ACO-OFDM component. For the $k$-th channel in (\ref{YHXZ6-7}) where $k$ is odd, for transmission at rate ${\cal R}_{1k}$, assume that a message $m$ is selected from ${{\cal M}} =\left\{1,...,\lfloor e^{\ell {\cal R}_{1k}}\rfloor\right\}$ uniformly randomly. The encoder maps $m$ to a length-$\ell$ codeword $\textbf{\texttt{X}}_{k}=\left[\texttt{X}_k^{(1)}(m),...,\texttt{X}_k^{(\ell)}(m)\right]^\textrm T$ in an ICG codebook ensemble.
In (\ref{YHXZ6-7}), $X_{1,k}$ and $D_{2,k}$ are independent since the ACO-OFDM component and the DCO-OFDM component are independent. So we can treat $D_{2,k}+Z_k$ as independent additive noise in the decoding of the ACO-OFDM component, and let the decoder follow a nearest neighbor decoding rule as
\setcounter{equation}{26}
\begin{equation}
\label{D1}
\hat{m}=\mathop{\arg\min}_{m\in{\cal M}}\frac{1}{\ell}\sum_{i=1}^\ell\left\|Y_{k}^{(i)}-\frac{1}{2}\texttt{X}_{k}^{(i)}(m)\right\|^2,\mspace{8mu} m\in {\cal M}.
\end{equation}
According to a result on the information rate of nearest neighbor decoding in additive non-Gaussian noise channels \cite{Lapidoth}, the information rate of the $k$th channel of the ACO-OFDM component is
${\cal R}_{1k}={\cal C}_\textrm{AWGN}\left(\gamma_2\right)$
where
$\gamma_2=\frac{\frac{1}{4}\sigma _{X1}^2}{\textrm E\left[D_{2,k}^2\right]+\sigma_z^2}$.
Since $\textrm E \left[|D_{2,k}|^2\right]=\textrm E \left[d_{2,n}^2\right]$ and $\sigma_{X1}^2=(1-\lambda)^2{4\pi}{\cal E}^2$, by noting that there are $\frac{N}{4}$ equivalent channel uses in frequency domain for the ACO-OFDM component (since there are $\frac{N}{4}$ odd numbered channels in (\ref{YHXZ6-7})) per $N$ channel uses in time domain, we get the information rate of the ACO-OFDM component with ICG codebook ensemble and nearest neighbor decoding as
\begin{equation}
\label{R1}
{\cal R}_1=\frac{N/4}{N}{\cal R}_{1k}=\frac{1}{4}\log\left(1+\frac{\pi(1-\lambda)^2{\cal E} ^2}{\textrm E\left[d_{2,n}^2\right]+\sigma_z^2}\right).
\end{equation}

For decoding the DCO-OFDM component, we first perform an interference cancellation procedure as follows. We reconstruct the ACO-OFDM component $\textbf{s}_1$ according to the decoding output of the nearest neighbor decoder given by (\ref{D1}), and subtract the reconstructed $\hat{\textbf{s}}_1$ from $\textbf{s}$. For the ACO-OFDM component, for any transmission rate below ${\cal R}_1$, the decoding error probability (i.e., the probability of $\hat{\textbf{s}}_1\neq\textbf{s}_1$) tends to zero as the code length $\ell$ grows without bound. So we can analyze the performance of the DCO-OFDM component based on
$\textbf{y}_2=\textbf{y}-\textbf{s}_1-{\cal A}=\textbf{c}_2+\textbf{z}$ where the DC bias has been removed, or $\textbf{Y}_2=\textbf{C}_2+\textbf{Z}$ in the frequency domain.
We rewrite the channel model as $\textbf{\textsf{Y}}_2=\textbf{\textsf{C}}_2+\textbf{\textsf{Z}}$,
where $\textbf{\textsf{Y}}_2=\left[Y_{2,2},Y_{2,4},...,Y_{2,\frac{N}{2}-2}\right]^\textrm T$, $\textbf{\textsf{C}}_2=\left[C_{2,2},C_{2,4},...,C_{2,\frac{N}{2}-2}\right]^\textrm T$, and $C_{2,k}=[\textbf{Fc}_2]_{2,k}$. Based on Appendix A, the GMI achieved by employing an ICG codebook ensemble and using a nearest neighbor decoding rule is
\begin{equation}
\label{GMI2}
{\cal I}_\textrm{GMI}=\left(\frac{N}{4}-1\right)\log\left(1+\frac{\Delta}{1-\Delta}\right)
\end{equation}
where
\begin{equation}
\Delta=\frac{\left|\textrm E\left[\textbf{\textsf{Y}}_2^\textrm H \textbf{\textsf{X}}_2\right]\right|^2}{\left(\frac{N}{4}-1\right)\sigma_X^2\left(\textrm E\left[\|\textbf{\textsf{C}}_2\|^2\right]+\textrm E\left[\|\textbf{\textsf{Z}}\|^2\right]\right)},
\end{equation}
where $\textbf{\textsf{X}}_2=\left[X_{2,2},X_{2,4},...,X_{2,\frac{N}{2}-2}\right]^\textrm T$.
Evaluating (\ref{GMI2}) by essentially the same steps as those in the proof of Theorem 1,
noting that $\textrm E \left[|D_{2,k}|^2\right]=\textrm E \left[d_{2,n}^2\right]$, and noting that there are $\frac{N}{4}-1$ equivalent channel uses in frequency domain for the DCO-OFDM component per $N$ channel uses in time domain, an achievable information rate of the DCO-OFDM component using ICG codebook ensemble and with nearest neighbor decoding is given by
\begin{equation}
\label{R2}
{\cal R}_2=\frac{\frac{N}{4}-1}{N}\log\left(1+\left(\textrm {erf}\left(\frac{\lambda{\cal E}}{\sqrt {2}\sigma _{x2}}\right)\right)^2\frac{2\sigma _{x2}^2}{\textrm E\left[d_{2,n}^2\right]+\sigma_z^2}\right).
\end{equation}

Combining (\ref{R1}) and (\ref{R2}), letting $\nu=\frac{1}{\sqrt 2\sigma_{X2}}$ and noting that $\sigma_{x2}^{2}=\frac{N-4}{2N}\sigma_{X2}^{2}$, letting $N\rightarrow\infty$, and jointly choosing $\lambda$ and $\nu$ that maximize the information rate, we obtain (\ref{ADO}).
\end{IEEEproof}

\begin{figure*}[b]
\begin{align}
\label{ACOl}
\textbf{X}_l=\bigg[&0,\textbf{0}^{2^{l-1}-1},X_{l,2^{l-1}},\textbf{0}^{2^{l}-1},X_{l,3\cdot2^{l-1}},...,\textbf{0}^{2^{l}-1},X_{l,\frac{N-2^{l}}{2}},\textbf{0}^{2^{l-1}-1}, \notag\\
&0,\textbf{0}^{2^{l-1}-1},\bar{X}_{l,\frac{N-2^{l}}{2}},\textbf{0}^{2^{l}-1},...,\bar{X}_{l,3\cdot2^{l-1}},\textbf{0}^{2^{l}-1},\bar{X}_{l,2^{l-1}},\textbf{0}^{2^{l-1}-1}\bigg]^\textrm T.
\tag{38}
\end{align}
\end{figure*}

\emph{Theorem 4}: The information rate of the HACO-OFDM with ICG and IG codebook ensembles for the ACO-OFDM component and the PAM-DMT component, respectively, and nearest neighbor decoding for both components, is
\begin{align}
\label{HACO}
{\cal R}
=\max_{0\leq\lambda\leq1}\bigg(&\frac{1}{4}\log\left(1+\frac{\pi(1-\lambda)^2{{\cal E}^2}}{\sigma_z^2}\right)\notag\\
&+\frac{1}{8}\log\left(1+\frac{2\pi\lambda^2{\cal E}^2}{\sigma_z^2}\right)\bigg).
\end{align}
The information rate of the ASCO-OFDM with ICG codebook ensemble
and nearest neighbor decoding for both components, is also given by (\ref{HACO}).

\begin{IEEEproof} We combine the proofs for these two schemes together since they share the same approach. An HACO-OFDM frame $\textbf{s}=\textbf{s}_1+\textbf{s}_2$ is the sum of an ACO-OFDM component and a PAM-DMT component which occupies only even subcarriers. A frame of the ASCO-OFDM input to the OIC as
$\textbf{s}=[\textbf{s}_1^\textrm T,\textbf{s}_2^\textrm T]^\textrm T=\left[[\textbf{s}_{\textrm A,1}+\textbf{s}_{\textrm{F,P}}]^\textrm T,[\textbf{s}_{\textrm A,2}+\textbf{s}_{\textrm{F,N}}]^\textrm T\right]^\textrm T$
is a length-$2N$ vector and is the sum of 1) two concatenated ACO-OFDM symbols denoted as $\textbf{s}_{\textrm A,1}$ and $\textbf{s}_{\textrm A,2}$, each of length $N$, and 2) a length-$2N$ frame of the Flip-OFDM component including two blocks, denoted as $[\textbf{s}_{\textrm{F,P}}^\textrm T, \textbf{s}_{\textrm{F,N}}^\textrm T]^\textrm T$, occuping only even subcarriers.
In both cases, the decoding of the ACO-OFDM component is not disturbed by the other component which generates clipping noise at only even subcarriers\footnote{When we consider only a single block in a frame of the Flip-OFDM component (for the ASCO-OFDM, this is needed when decoding the ACO-OFDM components), clipping noise also exists. It can be removed by combining two blocks of a frame of the Flip-OFDM.} \cite{RK14,WB15}, and therefore, the information rates of the ACO-OFDM components can be obtained by Theorem 2 directly.
Since the total transmit power is the sum of the powers of both components, we let the power of the ACO-OFDM component be $(1-\lambda) {\cal E}$, and the information rate of the ACO-OFDM component with ICG codebook ensemble and nearest neighbor decoding is thus
${\cal R}_1=\frac{1}{4}\log\left(1+\frac{\pi(1-\lambda)^2{\cal E}^2}{\sigma_z^2}\right)$.

Now consider the second component, which has a frequency domain input block as
\begin{align}
\textbf{X}_2=\big[&0,0,X_{2,2},0,X_{2,4},...,X_{2,\frac{N}{2}-2},0,\notag\\
&0,0,\bar{X}_{2,\frac{N}{2}-2},...,\bar{X}_{2,4},0,\bar{X}_{2,2},0\big]^\textrm T
\end{align}
for both schemes, but for the HACO-OFDM we have a further constraint of $\textrm {Re}[X_{2k}]\equiv0$.
For even $k$ satisfying $1\leq k \leq \frac{N}{2}-1$ we let $\textrm {Im}[X_{2k}]$ be i.i.d. and $\textrm {Im}[X_k]\sim {\cal N}\left(0,\sigma_{X2}^{2}\right)$ for the HACO-OFDM, and let $X_{2k}$ be i.i.d. and $X_k\sim {\cal{CN}}\left(0,\sigma_{X2}^{2}\right)$ for the ASCO-OFDM.
Taking IDFT of $\textbf{X}_2$, we obtain $\textbf{x}_2$ satisfying $\|\textbf{x}_2\|=\|\textbf{X}_2\|$ and $x_{2,n}\sim {\cal N}\left(0,\sigma_{x2}^{2}\right)$, where
$\sigma_{x2}^{2}=\frac{N-2}{2N}\sigma_{X2}^{2}$.
For both schemes, each element of $\textbf{s}_2$ satisfies $s_{2,n}\sim{\cal{TG}}\left(0,\sigma_{x2}^2\right)$
and the power it costs is
$\textrm E\left[s_{2,n}\right]=\frac{\sigma_{x2}}{\sqrt{2\pi}}$.
Letting $\textrm E\left[s_{2,n}\right] =\lambda{\cal E}$, the $\sigma_X$-$\cal E$ relationship is thus
${\sigma_{X2}^2}=\frac{2N}{N-2}\lambda^2{2\pi}{\cal E}^2$, for both schemes.

At the receiver, we first perform decoding of the ACO-OFDM component and the clipping noise cancellation process like that described in the proof of Theorem 3 to remove the clipping noise from the ACO-OFDM component.
The second component is then obtained which is clipping-noise-free, where the Flip-OFDM component of the ASCO-OFDM as $\textbf{r}_{\textrm{F}}=\big[[\textbf{s}_{\textrm{F,P}}+\textbf{n}_{1}]^\textrm T,[\textbf{s}_{\textrm{F,N}}+\textbf{n}_{2}]^\textrm T\big]^\textrm T$ is further transformed by the corresponding $\textsf{R}$ given in Table III. By taking DFT, for both schemes we obtain $\frac{N}{4}-1$ parallel channels,
each of the same form as the corresponding one in Table III, and achieving an information rate of $\frac{1}{2}{\cal C}_\textrm{AWGN}\left(\frac{\sigma_X^2}{2\sigma_z^2}\right)$.
Combining these with the $\sigma_X$-$\cal E$ relationship, and letting $N\rightarrow\infty$,
the information rate of the second component can be obtained as
${\cal R}_2=\frac{1}{8}\log\left(1+\frac{2\pi\lambda^2{\cal E}^2}{\sigma_z^2}\right)$, for both schemes.
The information rate (\ref{HACO}) is thus obtained by choosing $\lambda$ that maximizes the sum of ${\cal R}_1$ and ${\cal R}_2$, for both schemes.
\end{IEEEproof}

\emph{Corollary 1}: At high SNR, the asymptotically optimal power allocation parameters for the HACO-OFDM and the ASCO-OFDM are both
$\lim_{{\cal E}\rightarrow\infty}\lambda^*=\frac{1}{3}$. The corresponding asymptotic information rates are both
\begin{equation}
{\cal R}_\textrm{ASCO-OFDM}\sim\frac{3}{8}\log\left(\frac{\pi 2^{\frac{5}{3}}}{9}\frac{{\cal E}^2}{\sigma_z^2}\right).
\end{equation}
\begin{IEEEproof}
Since for $\frac{\cal E}{\sigma_z}\gg1$,
\begin{align}
\label{ASCO1}
&\frac{1}{4}\log\left(1+\frac{\pi(1-\lambda)^2{\cal E}^2}{\sigma_z^2}\right)+\frac{1}{8}\log\left(1+\frac{2\pi\lambda^2{\cal E}^2}{\sigma_z^2}\right)\notag\\
=&\frac{1}{4}\log\left(\left(1+\frac{\pi(1-\lambda)^2{\cal E}^2}{\sigma_z^2}\right)\left(1+\frac{2\pi\lambda^2{\cal E}^2}{\sigma_z^2}\right)^{\frac{1}{2}}\right)\notag\\
\sim &\frac{1}{4}\log\frac{\pi\sqrt{2\pi}\lambda(1-\lambda)^2{\cal E}^3}{\sigma_z^3},
\end{align}
the RHS of (\ref{ASCO1}) is maximized when $\lambda(1-\lambda)^2$ is maximized, which implies the optimal choice to be $\lim_{{\cal E}\rightarrow\infty}\lambda^*=\frac{1}{3}$.
The corresponding asymptotic information rates can be obtained straightforwardly.
\end{IEEEproof}

\section{Multi-Component Schemes}

This section studies the information rates of two multi-component unipolar OFDM schemes in Table I, namely, the FDM-UOFDM and the eU-OFDM. Our results show that with ICG codebook ensemble and nearest neighbor decoding, they have the same information rate, which closely approaches the high-SNR capacity of the Gaussian OIC under average power constraint.

\emph{Theorem 5}: The information rate of the FDM-UOFDM with $L$ ACO-OFDM components, each employing ICG codebook ensemble and nearest neighbor decoding, is
\begin{align}
\label{SEE}
&{\cal R}_{L\textrm{-FDM-UOFDM}}\notag\\
&=\max_{\substack{\lambda_1,...,\lambda_L:\\ \lambda_l\ge0, \sum_{l=1}^L\lambda_l=1}}\sum_{l=1}^L\frac{1}{2^{l+1}}\log\left(1+\frac{2^{l-1}\pi\lambda_l^2{\cal E}^2}{\sigma_z^2}\right).
\end{align}

\begin{IEEEproof}Let $N$ be divisible by $2^L$ where $L\leq\log_2{N}-1$. The input of the first ACO-OFDM component of the FDM-UOFDM is
\begin{align}
\textbf{X}_1= \big[&0,X_{1,1},0,X_{1,3}, ...,0,X_{1,\frac{N}{2}-1}, \notag\\
&0, \bar{X}_{1,\frac{N}{2}-1},0,...,\bar{X}_{1,3},0,\bar{X}_{1,1}\big]^\textrm T,
\end{align}
and in general the $l$-th ACO-OFDM component of the FDM-UOFDM is given by (\ref{ACOl})
where $\textbf{0}_n$ denotes a length-$n$ all-zero vector.
Apparently, each subcarrier is occupied by exactly one component before asymmetric clipping, except for subcarriers $0$ and $\frac{N}{2}$. Let $X_{l,k}$ be i.i.d. and $X_{l,k}\sim{\cal {CN}}\left(0,\sigma_{Xl}^2\right)$ for $k$ belonging to
\setcounter{equation}{38}
\begin{equation}
\label{kset}
{\cal K}=\left\{k: k=(2m-1)2^{l-1}, 1\leq m \leq \frac{N}{2^{l+1}}\right\}.
\end{equation}
Taking IDFT of $\textbf{X}_l$, we obtain $\textbf{x}_l$ satisfying $\|\textbf{x}_l\|=\|\textbf{X}_l\|$ and $x_{l,n}\sim {\cal N}\left(0,\sigma_{xl}^{2}\right)$, where
$\sigma_{xl}^{2}=2^{-l}\sigma_{Xl}^{2}$.
A frame of the unipolar FDM-UOFDM input to the OIC, denoted as $\textbf{s}=[s_1,...,s_{N-1}]^\textrm T$, is the superposition of all ACO-OFDM components as
$\textbf{s}=\sum_{l=1}^L \textbf{s}_l$
where $\textbf{s}_l$ is obtained by asymmetrically clipping $\textbf{x}_l$. Since the ACO-OFDM components are mutually independent we have
$\textrm E\left[s_n\right]=\sum_{l=1}^L \textrm E\left[s_{l,n}\right]$
where $s_{l,n}\sim{\cal{TG}}\left(0,\sigma_{xl}^2\right)$,
and the power it costs is
$\textrm E\left[s_{l,n}\right]=\frac{\sigma_{xl}}{\sqrt{2\pi}}$.
Letting $\textrm E\left[s_{l,n}\right]=\lambda_l{\cal E}$, the $\sigma_X$-$\cal E$ relationship is thus
$\sigma_{Xl}^2=2^l\lambda_l^2{2\pi}{\cal E}^2$
where $\{\lambda_l\}$ satisfy
$\sum_{l=1}^{L}\lambda_l\leq1$.

Note that the clipping noise of the $l$-th component of the FDM-UOFDM falls on the subcarriers occupied by the $(l+1)$-th component and subcarriers $0$ and $\frac{N}{2}$. At the receiver, the decoding of the first ACO-OFDM component is the same as that described in the proof of Theorem 2. So the information rate
${\cal R}_{1}=\frac{1}{4}\log\left(1+\frac{\pi\lambda_1^2{\cal E}^2}{\sigma_z^2}\right)$
can be achieved by ICG codebook ensemble and nearest neighbor decoding. When the transmission rate of the first ACO-OFDM component is below ${\cal R}_{1}$, we can perform the clipping noise cancellation procedure like that described in the proof of Theorem 3 so that the clipping noise from the first ACO-OFDM component is removed and the decoding of the second ACO-OFDM component is clipping-noise-free. This procedure can be performed recursively so that the decoding of each ACO-OFDM component is also clipping-noise-free. We can obtain $2^{-(l+1)}N$ parallel channels for the $l$-th ACO-OFDM component as
$Y_{l,k}=\frac{1}{2} X_{l,k}+Z_k$,
where $k$ belongs to the set $\cal K$.
The information rate for each $k$
when $X_{l,k} \sim{\cal {CN}}\left(0,\sigma_{Xl}^2\right)$ is $C_\textrm{AWGN}\left(\frac{\sigma_{Xl}^2}{4\sigma_z^2}\right)$.
The information rate of this component with ICG codebook ensemble and nearest neighbor decoding is thus
${\cal R}_{l}=2^{-(l+1)}\cdot C_\textrm{AWGN}\left(\frac{\sigma_{Xl}^2}{4\sigma_z^2}\right)$.
Combining the $\sigma_X$-$\cal E$ relationship, this yields
\begin{equation}
\label{SEEl}
{\cal R}_{l}=2^{-(l+1)}\log\left(1+\frac{2^{l-1}\pi\lambda_l^2{\cal E}^2}{\sigma_z^2}\right).
\end{equation}
The proof is completed by summing up $\{{\cal R}_{l}\}$ over $l$ and choosing $\{\lambda_l\}$ maximizing the information rate.
\end{IEEEproof}

\emph{Theorem 6}: The information rate of the eU-OFDM with $L$ components, each employing ICG codebook ensemble and nearest neighbor decoding, is
\begin{align}
\label{eU}
&{\cal R}_{L\textrm{-eU-OFDM}}\notag\\
&=\max_{\substack{\lambda_1,...,\lambda_L:\\ \lambda_l\ge0, \sum_{l=1}^L\lambda_l=1}}\sum_{l=1}^L\frac{1}{2^{l+1}}\log\left(1+\frac{2^{l-1}\pi\lambda_l^2{\cal E}^2}{\sigma_z^2}\right).
\end{align}
\begin{figure*}
\centering
\includegraphics[width=6in,height=4.5in]{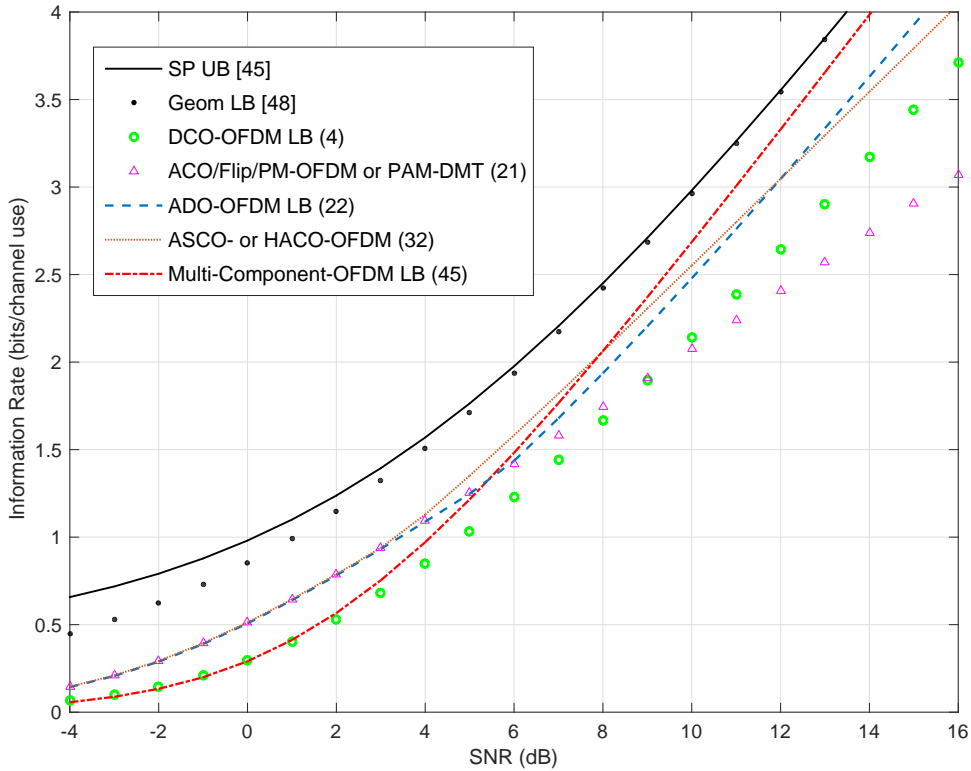}
\caption{Results on information rates of unipolar OFDM in discrete-time Gaussian OIC under average power constraint.}
\label{OOFDM}
\end{figure*}

\begin{IEEEproof}
Consider the case that the Flip-OFDM components are used.
Let the $m$-th block of the input of the $l$-th Flip-OFDM component be
\begin{equation}
\textbf{X}_{l,m}=\left[0,X_{l,m,1},...,X_{l,m,\frac{N}{2}-1},0,\bar{X}_{l,m,\frac{N}{2}-1},...,\bar{X}_{l,m,1}\right]^\textrm T
\end{equation}
which is a length-$N$ complex vector with Hermitian symmetry. For $1\leq k \leq \frac{N}{2}-1$, let $X_{l,m,k}$ be i.i.d. and $X_{l,m,k}\sim {\cal{CN}}\left(0,\sigma_{Xl}^{2}\right)$. Taking DFT of $\textbf{X}_{l,m}$ we obtain $\textbf{x}_{l,m}$ satisfying $\|\textbf{x}_{l,m}\|=\|\textbf{X}_{l,m}\|$ and $x_{l,n}\sim {\cal{N}}\left(0,\sigma_{xl}^{2}\right)$, where
$\sigma_{xl}^{2}=\frac{N-2}{N}\sigma_{Xl}^{2}$.
Denote a corresponding Flip-OFDM symbol as
$\left[\textbf{s}_{l,m,\textrm{P}}^{\textrm T},\textbf{s}_{l,m,\textrm{N}}^{\textrm T}\right]^{\textrm T}$.
The $l$-th Flip-OFDM component of a frame of the eU-OFDM is
\begin{align}
\textbf{s}_l=\bigg[&\left[\textbf{1}_{2^{l-1}}\otimes\textbf{s}_{l,1,\textrm{P}}\right]^{\textrm T},\left[\textbf{1}_{2^{l-1}}\otimes\textbf{s}_{l,1,\textrm{N}}\right]^{\textrm T},\notag\\
&...,\left[\textbf{1}_{2^{l-1}}\otimes\textbf{s}_{l,2^{L-l},\textrm{P}}\right]^{\textrm T},\left[\textbf{1}_{2^{l-1}}\otimes\textbf{s}_{l,2^{L-l},\textrm{N}}\right]^{\textrm T}\bigg]^{\textrm T}
\end{align}
where $\otimes$ denotes Kronecker product and $\textbf{1}_l$ denotes a length-$l$ all-one vector. The length of such a frame is $2^LN$, which is determined by the $L$-th component. A frame of the eU-OFDM is thus
$\textbf{s}=\sum_{l=1}^L \textbf{s}_l$.
Since the Flip-OFDM components are mutually independent we have
$\textrm E\left[s_n\right]=\sum_{l=1}^L \textrm E\left[s_{l,n}\right]$
where $s_{l,n}\sim{\cal{TG}}\left(0,\sigma_{xl}^2\right)$
and the power it costs is thus
$\textrm E\left[s_{l,n}\right]=\frac{\sigma_{xl}}{\sqrt{2\pi}}$.
We let $\textrm E\left[s_{l,n}\right]=\lambda_l{\cal E}$, i.e., $\sigma_{xl}=\lambda_l\sqrt{2\pi}{\cal E}$,
where $\{\lambda_l\}$ satisfy
$\sum_{l=1}^{L}\lambda_l=1$.

At the receiver the first Flip-OFDM component can be decoded as that described in Table III
because the receive transform $\textsf{R}$ also cancels all other components since they are spreaded by the all-one vector. The information rate of the first Flip-OFDM component with ICG codebook ensemble and nearest neighbor decoding is thus
${\cal R}_{1}=\frac{1}{4}\log\left(1+\frac{\pi\lambda_1^2{\cal E}^2}{\sigma_z^2}\right)$.
When the transmission rate of the first Flip-OFDM component is below ${\cal R}_{1}$, we can perform the clipping noise cancellation procedure like that described in the proof of Theorem 3 so that the clipping noise from the first Flip-OFDM component can be removed and the decoding of the second Flip-OFDM component is clipping-noise-free. This procedure can be performed recursively so that the decoding of each Flip-OFDM component is also clipping-noise-free. Moreover, the decoding of the $l$-th Flip-OFDM component includes a despreading process which
introduces a spreading gain of $2^{l-1}$ on the SNR of the equivalent channel, and a
rate loss factor of $2^{-(l-1)}$. So the information rate of the $l$-th Flip-OFDM component with ICG codebook ensemble and nearest neighbor decoding is
\begin{equation}
\label{eU2}
{\cal R}_{l}=\frac{1}{2^{l+1}}\log\left(1+\frac{2^{l-1}\pi\lambda_l^2{\cal E}^2}{\sigma_z^2}\right)
\end{equation}
and the proof is completed by summing up $\{{\cal R}_{l}\}$ over $l$ and choosing $\{\lambda_l\}$ maximizing the information rate.
\end{IEEEproof}

\emph{Corollary 2}: The asymptotic information rates of the $L$-component FDM-UOFDM and the eU-OFDM as $L\rightarrow\infty$, with ICG codebook ensemble and nearest neighbor decoding for each component, are both lower bounded by
$\frac{1}{2}\log\left(\frac{\pi}{8}\frac{{\cal E}^2}{\sigma_z^2}\right)$.

\begin{IEEEproof}
By letting
$\lambda_l=2^{-l}$ (i.e., $\sigma_{Xl}^2=2^{-l}{2\pi}{\cal E}^2$,
noting that $\sum_{l=1}^L 2^{-l}<1$), the RHS of (\ref{SEE}) or (\ref{eU}) can be lower bounded by
\begin{align}
\label{SEE2}
&\lim_{L\rightarrow\infty}\sum_{l=1}^L\frac{1}{2^{l+1}}\log\left(1+\frac{\pi{\cal E}^2}{2^{l+1}\sigma_z^2}\right)\notag\\
>&\lim_{L\rightarrow\infty}\frac{1}{2}\log\left(\left(2^{-(l+1)}\frac{\pi{\cal E}^2}{\sigma_z^2}\right)^{\sum_{l=1}^L{2^{-l}}}\right)\notag\\
=&\frac{1}{2}\log\left(\lim_{L\rightarrow\infty}2^{\sum_{l=1}^L\frac{-(l+1)}{2^{l}}} \left(\frac{\pi{\cal E}^2}{\sigma_z^2}\right)^{\sum_{l=1}^L{2^{-l}}}\right)\notag\\
=&\frac{1}{2}\log\left(2^{\lim_{L\rightarrow\infty}\sum_{l=1}^L\frac{-(l+1)}{2^{l}}} \left(\frac{\pi{\cal E}^2}{\sigma_z^2}\right)\right)\notag\\
=&\frac{1}{2}\log \left(\frac{\pi}{8}\frac{{\cal E}^2}{\sigma_z^2}\right).
\end{align}
\end{IEEEproof}

In the proof of Corollary 2 we use an unequal power allocation strategy
By contrast, it can be readily verified that the information rates of multi-component schemes obtained by equal power allocation (i.e., $\lambda_l=1/L$ for all $1\leq l\leq L$) tends to zero as $L$ tends to infinity.

The capacity of the discrete-time Gaussian OIC satisfies \cite{HK04}, \cite{LMW09}
\begin{equation}
\label{bounds}
\frac{1}{2}\log\left(1+\frac{e}{2\pi}\frac{{\cal E}^2}{\sigma_z^2}\right)\leq {\cal C}_\textrm{DTOIC}\leq\frac{1}{2}\log\left(\frac{e}{2\pi}\left(\frac{{\cal E}}{\sigma_z}+2\right)^2\right).
\end{equation}
According to Corollary 2, at high SNR the gap between the information rate of multi-component schemes and the capacity of the discrete-time Gaussian OIC is at most
$10\log_{10}\frac{2\sqrt{e}}{\pi}\thickapprox0.21 \mspace{4mu} \textrm{dB}$.
In other words, multi-component schemes can approach the high-SNR capacity of the discrete-time Gaussian OIC to within 0.07 bits.

\section{Numerical Results and Discussions}

Fig. \ref{OOFDM} shows our main results on the information rates of unipolar OFDM schemes in the Gaussian OIC under average power constraint, where $\textrm{SNR}\triangleq\frac{\cal E}{\sigma_z}$ is the optical SNR. The sphere packing based upper bound (SP UB) and the geometrically distributed input based lower bound (Geom LB) on the capacity of the Gaussian OIC are shown as benchmarks, which bound the channel capacity to within a small gap. See (\ref{bounds}) for their expressions.
At low SNR, it is shown that single-component schemes including the ACO-OFDM, the PAM-DMT, the Flip-OFDM, and the PM-OFDM, and double-component schemes including the ADO- and the ASCO-OFDM, have the same information rate which is higher than information rates of other schemes. However, all achievability results still have considerable gaps to the low-SNR capacity. At high SNR double-component schemes achieves higher information rates than single-component ones, and multi-component schemes including the FDM-UOFDM and the eU-OFDM with optimized power allocation, closely approach the capacity of the Gaussian OIC asymptotically.

\begin{figure}
\centering
\includegraphics[width=3.48in,height=2.61in]{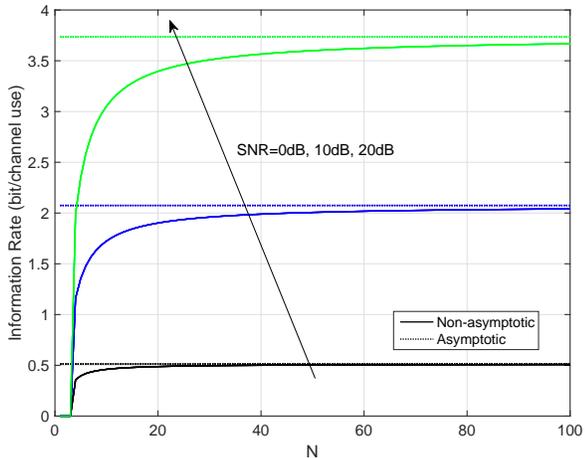}
\caption{Information rates of ACO-OFDM with respect to the number of subcarriers.}
\label{Asy}
\end{figure}

In addition, since our results are asymptotic ones for large $N$,
we compare our results and non-asymptotic results in Fig. \ref{Asy},
where the ACO-OFDM is taken as an example.
It is shown that our results maintain high accuracy for practical values of $N$, e.g., $N\ge64$.

\subsection{DCO-OFDM Parameter Optimization}

\begin{figure}
\begin{minipage}[t]{1\linewidth}
\centering
\includegraphics[width=3.48in,height=2.61in]{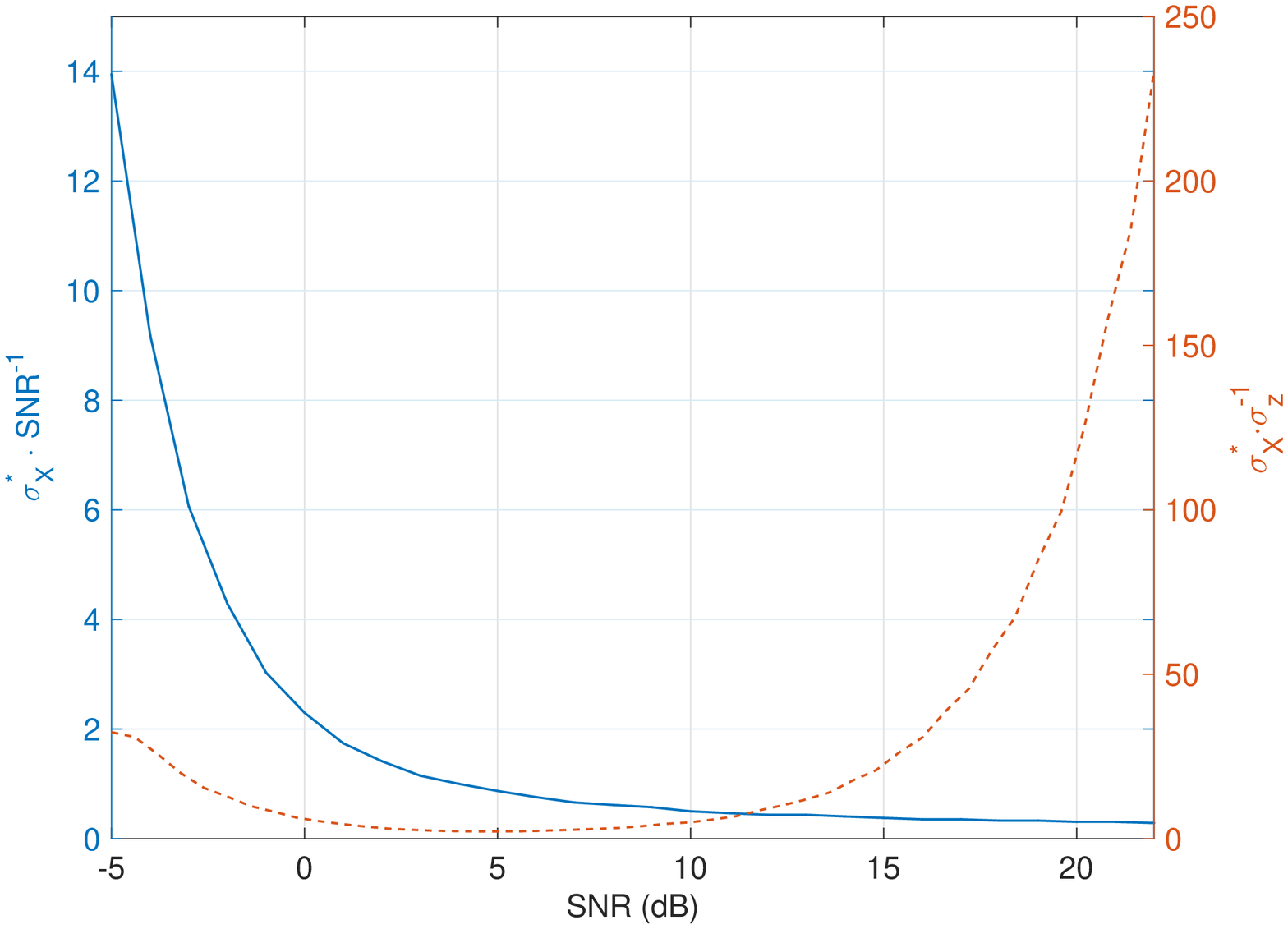}
\caption{Optimal $\sigma_X$ maximizing information rate of DCO-OFDM.}
\label{DCO0}
\end{minipage}
\begin{minipage}[t]{1\linewidth}
\centering
\includegraphics[width=3.48in,height=2.61in]{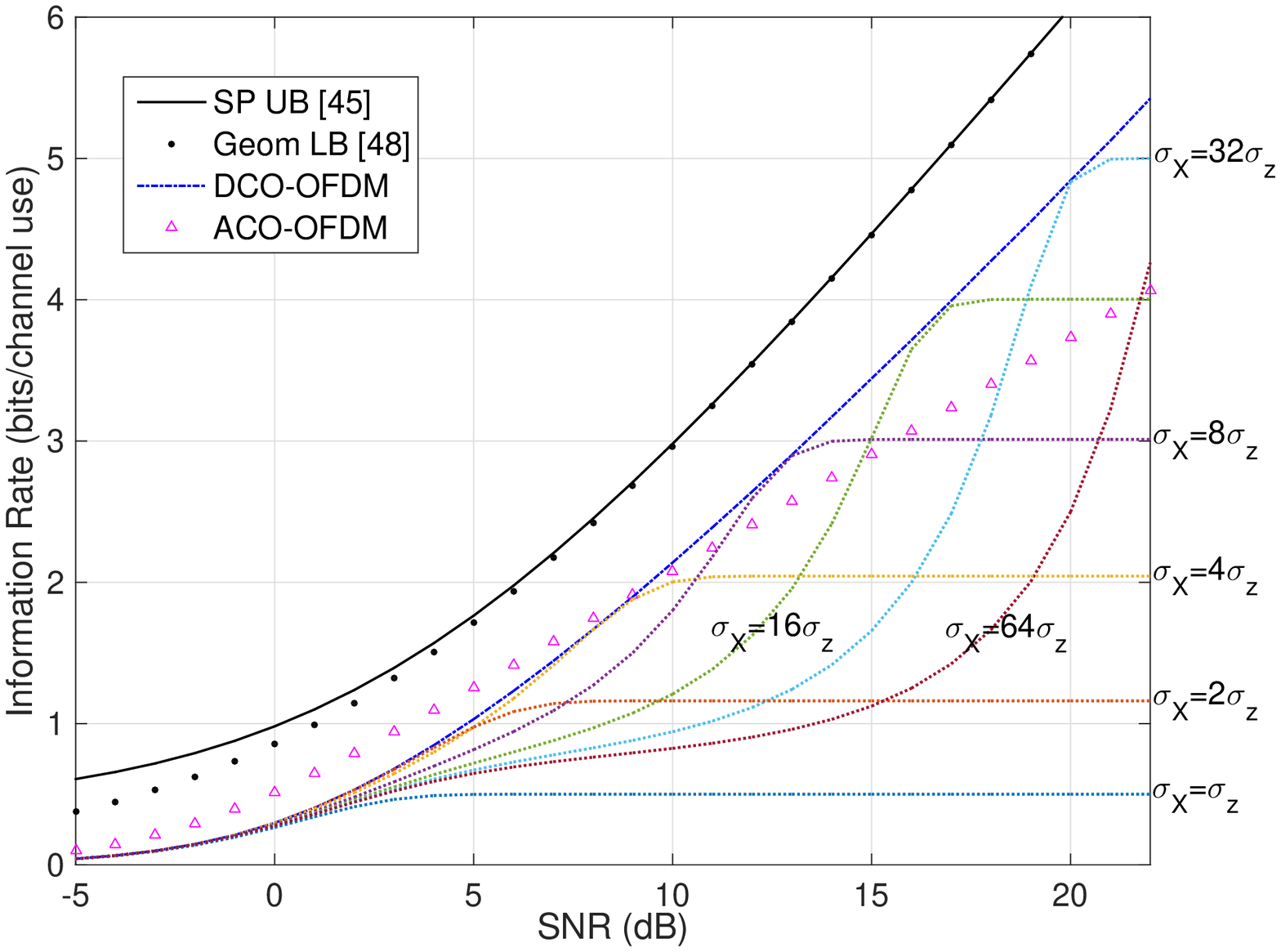}
\caption{Information rates of DCO-OFDM for $\sigma_X\propto\sigma_z$.}
\label{DCO1}
\end{minipage}
\begin{minipage}[t]{1\linewidth}
\centering
\includegraphics[width=3.48in,height=2.61in]{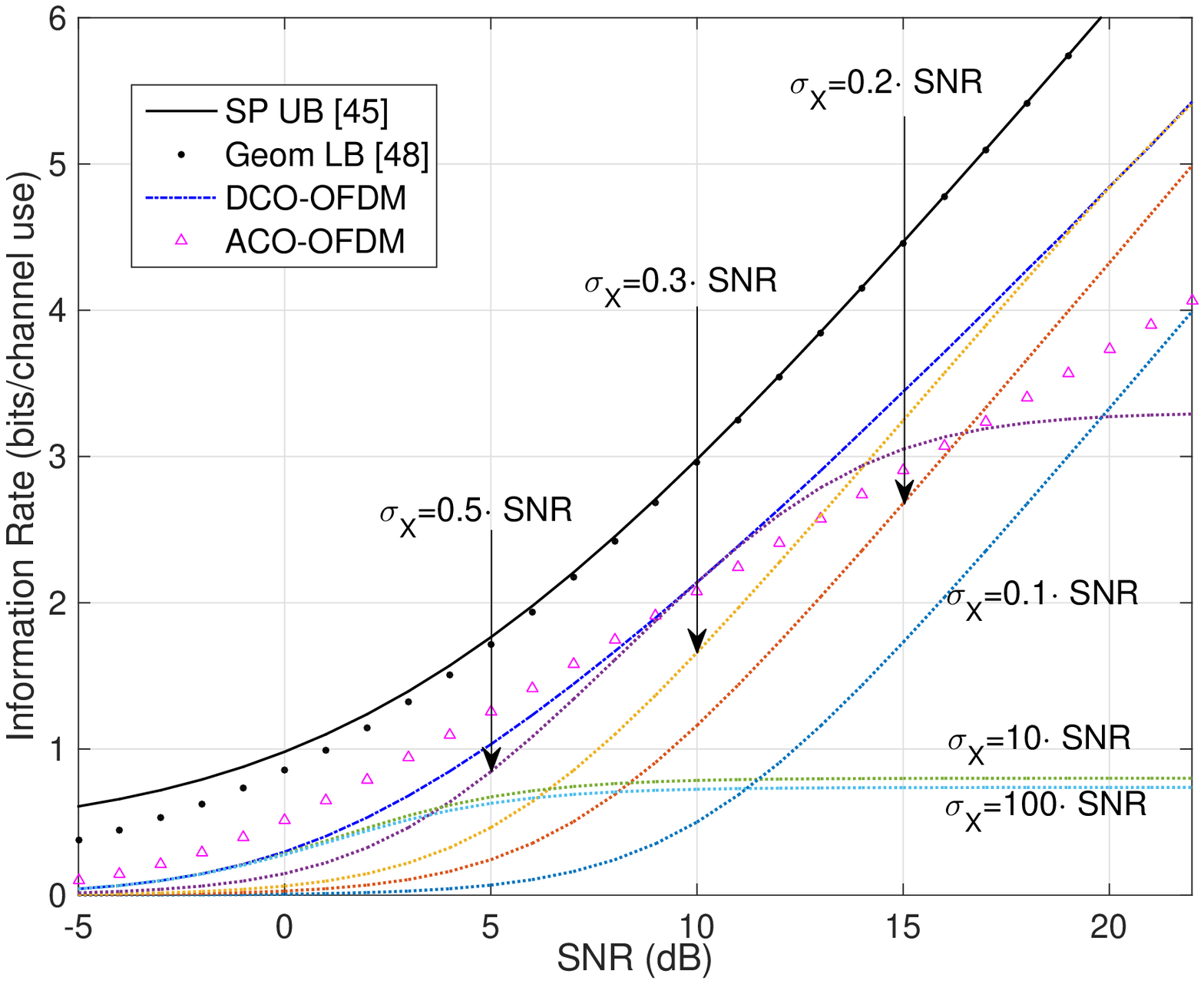}
\caption{Information rates of DCO-OFDM for $\sigma_X\propto\textrm{SNR}$.}
\label{DCO2}
\end{minipage}
\end{figure}

The comparison of the DCO- and the ACO-OFDM have been widely investigated, e.g., \cite{JS09,MEH11,DA13}.
Fig. \ref{OOFDM} shows that when SNR is below 9 dB the ACO-OFDM performs better, otherwise the DCO-OFDM performs better. However, the optimal value of $\sigma_X$ (or $\nu$, equivalently) in the DCO-OFDM varies with SNR, which renders challenges for the design of practical systems. The problem of optimizing parameters for the DCO-OFDM has also been widely studied \cite{DSH121,DH13,SEUZhang,SEUZhao}, where different constraints and performance metrics were used. For example, \cite{DH13} proposed a parameter optimization framework for both the DCO-OFDM and the ACO-OFDM (clipped from above) with respect to information rate, under electrical and optical power constraints, as well as dynamic range constraints.
Fig. \ref{DCO0} shows the optimal $\sigma_X$, denoted as $\sigma_X^*$, that maximizes the information rate of the DCO-OFDM under our assumptions, . Obviously, neither a fixed $\sigma_X$ nor a $\sigma_X$ being proportional to SNR (with a fixed ratio) can achieve the best performance at low and high SNR simultaneously.
In Fig. \ref{DCO1}, it is shown that when $\sigma_X$ is fixed, using a relatively small $\sigma_X$ approaches the best performance of the DCO-OFDM when $\textrm{SNR}< 10$ dB. At high SNR a larger $\sigma_X$ must be used. However, in this case a fixed $\sigma_X$ causes performance loss which becomes larger as SNR increases.
In Fig. \ref{DCO2}, $\sigma_X$ is increased as SNR increases. It is shown again that a fixed ratio between $\sigma_X$ and SNR cannot achieve the best performance at low and high SNR simultaneously. Note that for given $\textrm{SNR}$, a larger $\sigma_X$ causes a larger clipping noise. However, when the ratio $\frac{\sigma_X}{\cal E}$ tends to infinity (i.e., $\nu\rightarrow0$), the high-SNR information rate of the DCO-OFDM does not tend to zero although the probability that the IDFT output $\textbf{x}$ is clipped tends to one. In fact, we can show that
\begin{align}
\lim_{\nu\rightarrow0, \mspace{4mu}{\cal E}\rightarrow\infty} {\cal R}_\textrm{DCO-OFDM}&=\frac{1}{2}\log\frac{\pi}{\pi-2}\notag\\
&\approx0.73 \mspace{8mu}\textrm{bits/channel use}.
\end{align}
The interpretation of this fact is as follows: in this case the DCO-OFDM tends to a scheme employing IG codebook ensemble, nearest neighbor decoding, and \emph{binary output quantization} in Gaussian channel, so ${\cal R}_\textrm{DCO-OFDM}$ tends to the corresponding GMI which is 0.73 bits/channel use, as shown in [\ref{Boss}, Sec. IV].

\subsection{Power Allocation for Multiplexing Based Schemes}

\begin{figure}
\begin{minipage}[t]{1\linewidth}
\centering
\includegraphics[width=3.48in,height=2.61in]{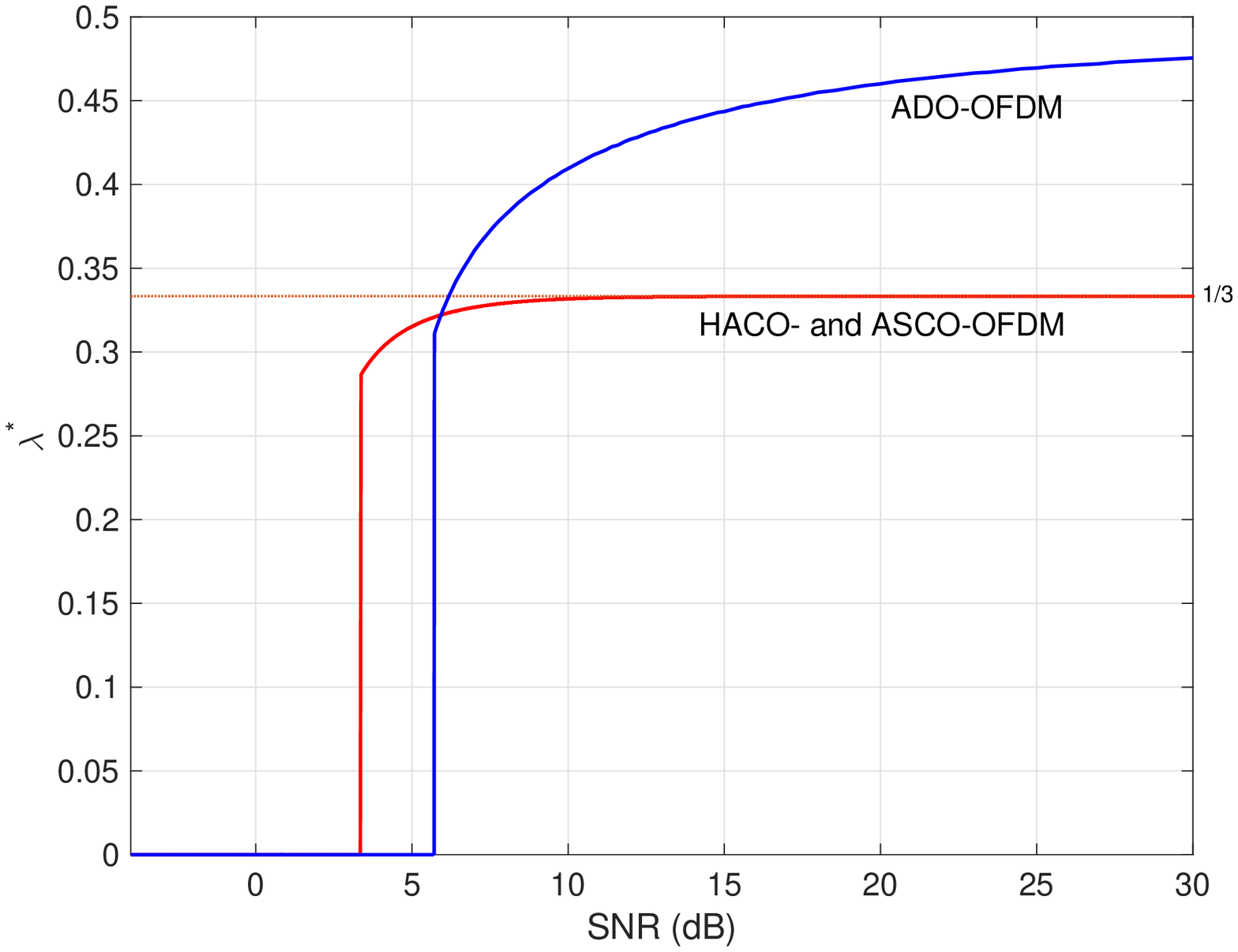}
\caption{Optimal power allocation parameters for double-component schemes.}
\label{ADOlambdaa}
\end{minipage}
\begin{minipage}[t]{1\linewidth}
\centering
\includegraphics[width=3.48in,height=2.61in]{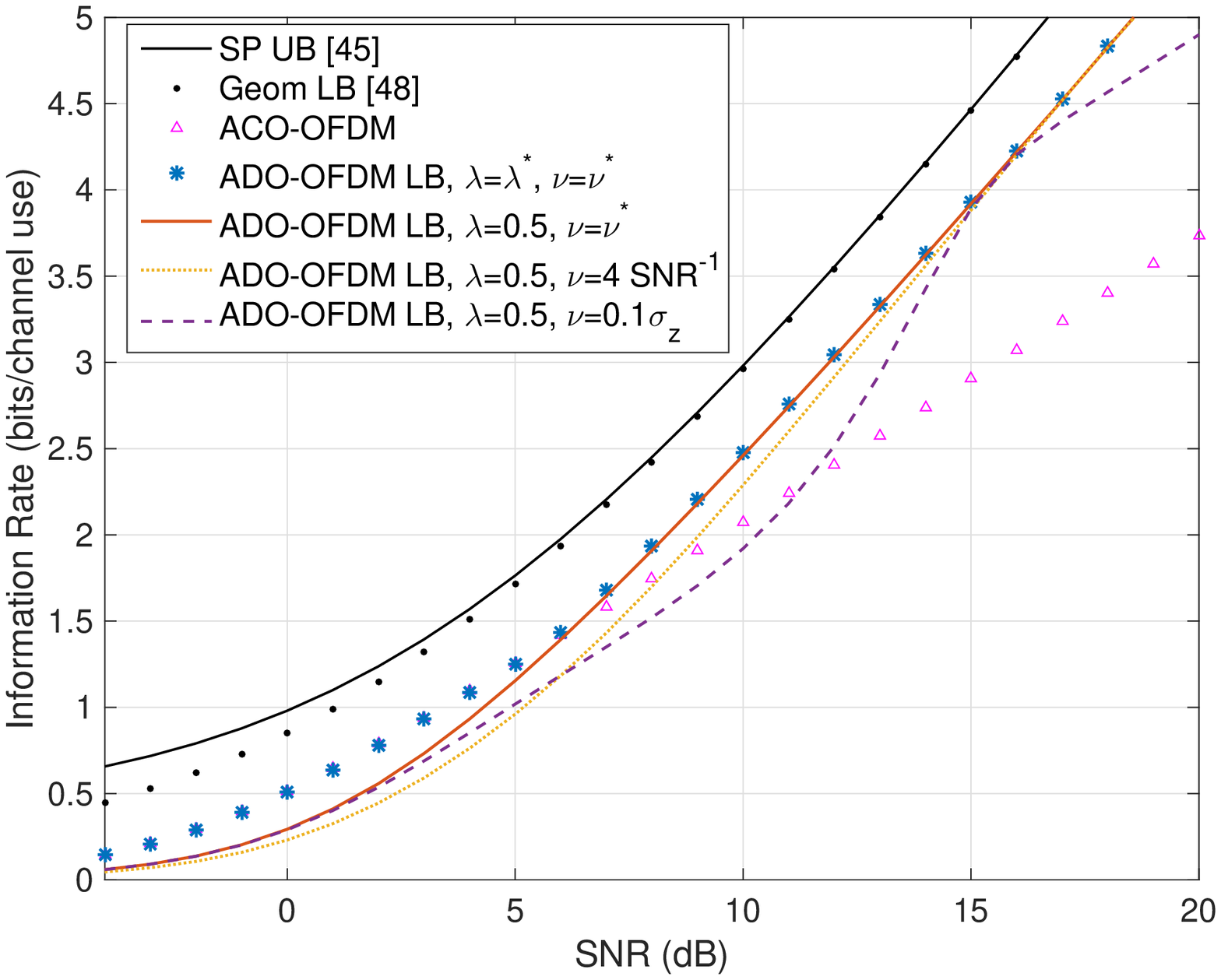}
\caption{Information rates of ADO-OFDM with and without optimal parameters.}
\label{adoopt}
\end{minipage}
\begin{minipage}[t]{1\linewidth}
\centering
\includegraphics[width=3.48in,height=2.61in]{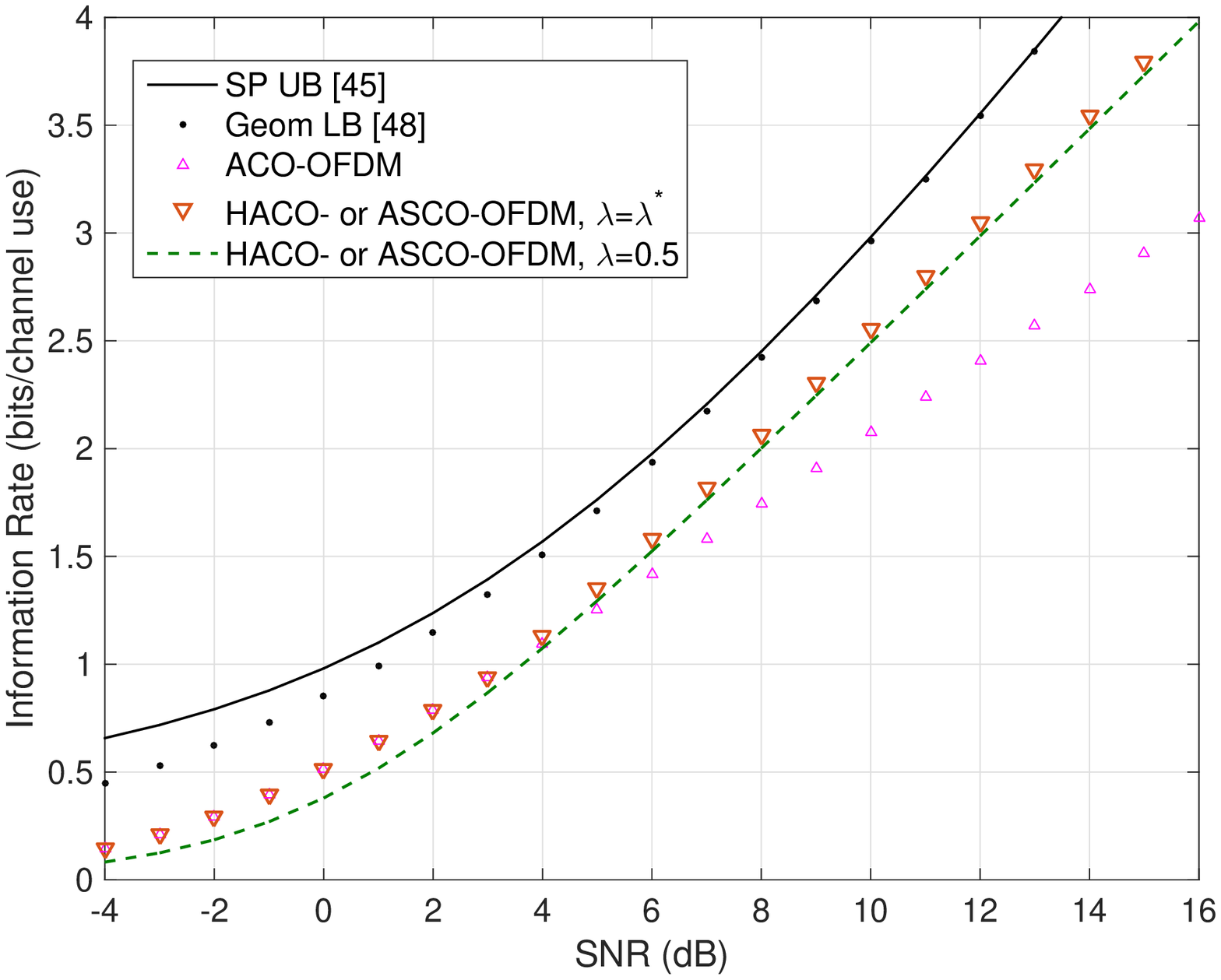}
\caption{Information rates of HACO- and ASCO-OFDM with and without optimal power allocation.}
\label{hacoopt}
\end{minipage}
\end{figure}

\begin{figure}
\begin{minipage}[t]{1\linewidth}
\centering
\includegraphics[width=3.48in,height=2.61in]{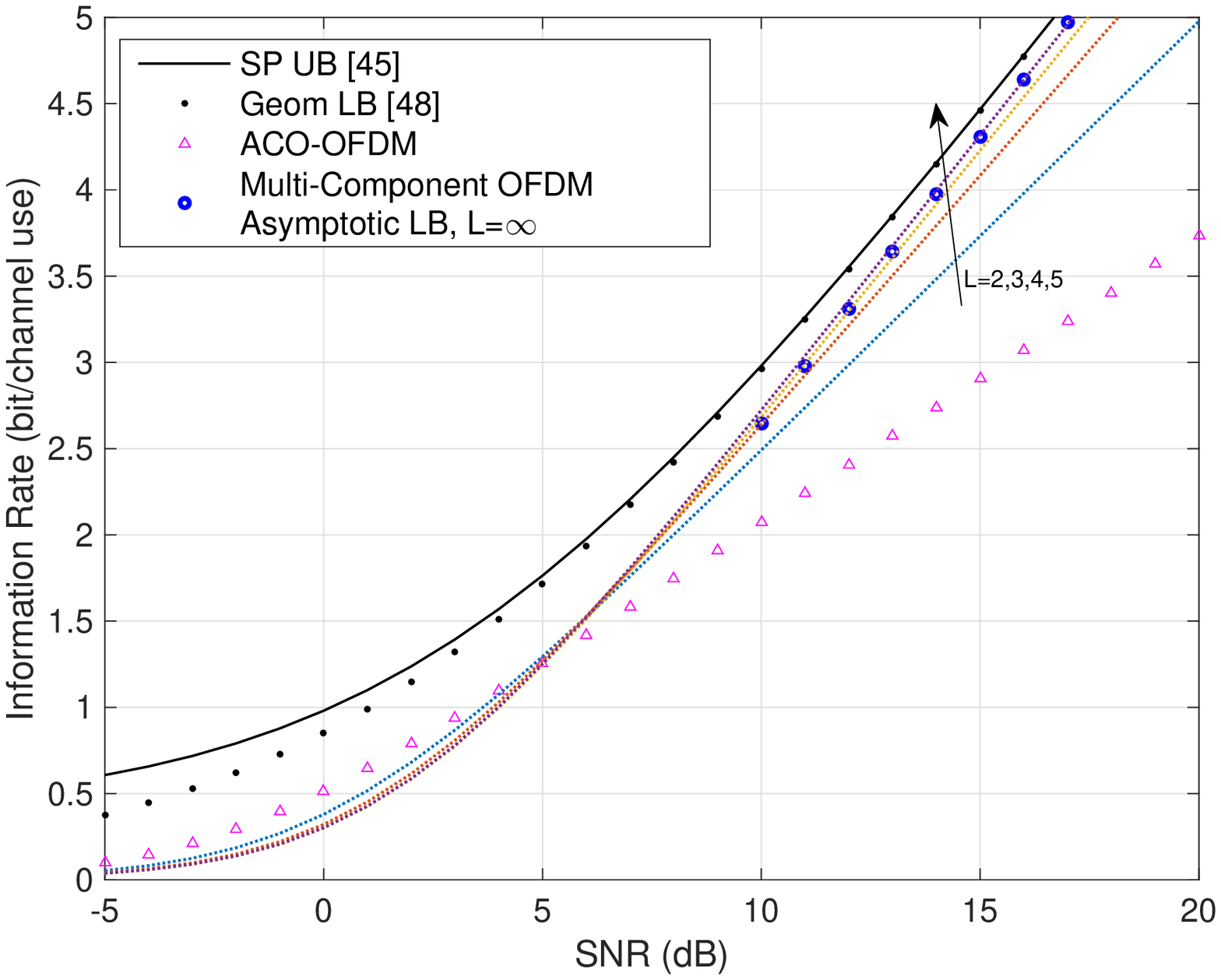}
\caption{Information rates of multi-component schemes with $l$ components.}
\label{Multi}
\end{minipage}
\begin{minipage}[t]{1\linewidth}
\centering
\includegraphics[width=3.48in,height=2.61in]{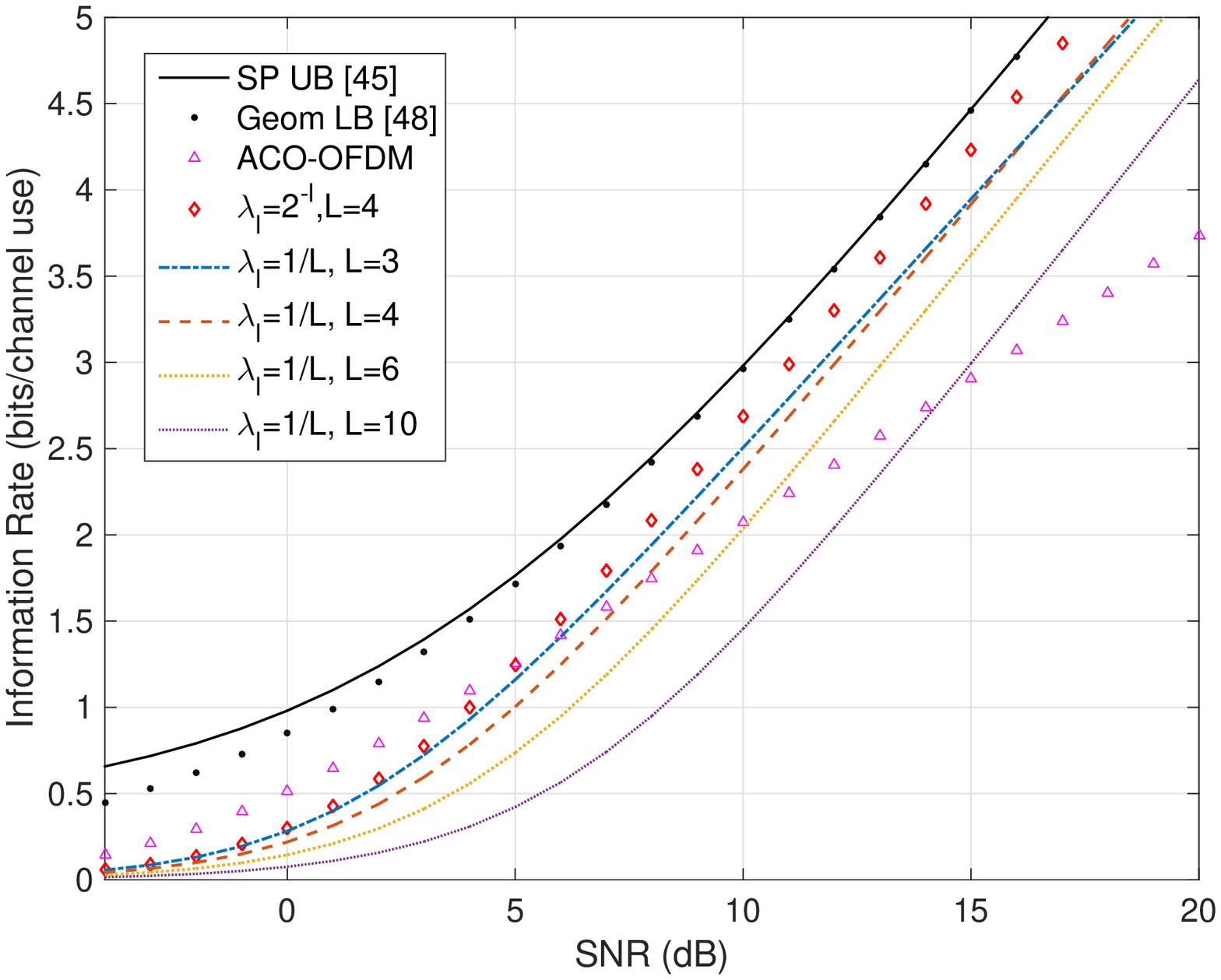}
\caption{Information rates of multi-component schemes with different power allocation strategies.}
\label{Multiopt}
\end{minipage}
\end{figure}

For multiplexing based schemes, the benefit of unequal power allocation among components has been shown in \cite{RK14}, \cite{SYG17}, which consider only uncoded transmission. This subsection provides our results on power allocation of multiplexing based schemes, in terms of information rates.

Fig. \ref{OOFDM} shows that double-component schemes achieve the same information rate as the ACO-OFDM at low SNR. This is because the optimal strategy then is allocating all power to the ACO-OFDM component.
Fig. \ref{ADOlambdaa} provides optimal power allocation parameters (i.e., $\lambda^*$) for double-component schemes. It is shown that both curves have a jump, which is at 5.71 dB and 3.36 dB for the ADO-OFDM and the HACO- or ASCO-OFDM, respectively.
For the HACO- and the ASCO-OFDM, the numerical results demonstrate the validity of Corollary 1, i.e. $\lim_{{\cal E}\rightarrow\infty}\lambda^*=\frac{1}{3}$. However, for the ADO-OFDM the asymptotically optimal choice of $\lambda$ at high SNR is unknown.
In Fig. \ref{adoopt}, it is shown that for ADO-OFDM, power allocation is extremely important at low SNR, and parameter optimization for the DCO-OFDM component is important at high SNR. For the HACO- or ASCO-OFDM, as shown in Fig. \ref{hacoopt}, the rate penalty due to suboptimal power allocation is smaller compared to the ADO-OFDM, but at low SNR the penalty is still considerable.

These results imply that we can introduce the following simple switching strategy in a double-component unipolar OFDM scheme, without significant performance loss. Take the ASCO-OFDM as an example. At low SNR, the Flip-OFDM component is inactive (i.e. no power is allocated) and the ASCO-OFDM reduces to the ACO-OFDM. When SNR is higher than a threshold, the Flip-OFDM component becomes active, and its power is allocated according to a fixed power allocation parameter (e.g., $\lambda=1/3$).

In Fig. \ref{Multi}, results on the information rates of multi-component schemes with $L$ components are given, where we set $\lambda_l=2^{-l}$ for $1\leq l\leq L-1$ and $\lambda_L=2^{-(L-1)}$. It is shown that a relatively small $L$ is already good enough. For example, multi-component schemes approach the high-SNR capacity of the Gaussian OIC to within 1 dB when $L$ is four.
The power allocation strategy $\lambda_l=2^{-l}$ is compared with equal power allocation in Fig. \ref{Multiopt}, which shows that employing four components with the former strategy, the achieved information rate outperforms that achieved using equal power allocation (it decreases when $L$ exceeds four).
\section{Concluding Remarks}

In this paper, we derive an array of information rate results for unipolar OFDM schemes in average power constrained OIC. The results validate existing SNDR based analysis, and provide further insights on transceiver design.
The equivalence of information rates of several different schemes is established. Moreover, we demonstrate the benefit of component multiplexing for unipolar OFDM, find near-optimal power allocation strategies, and show that several multi-component unipolar OFDM schemes are near-optimal at high SNR.

This work can be extended to Gaussian OIC under other types of input constraints (see these listed in Sec. II), or Gaussian OIC with time dispersion (frequency selectivity), note that the obtained results may be quite different from ours.
When a peak power constraint exists, for unipolar OFDM schemes with unbounded channel input,
their peaks must be clipped (or reduced by a nonlinear transform), and their information rates can also be lower bounded by evaluating the GMI. For multiplexing based schemes the clipping can be performed in each component, and the information rate derivation follows the approach for the ADO-OFDM. In these cases, all the schemes would require parameter optimization and optimized power allocation.

Moreover, our results can be extended to continuous-time Gaussian OICs, which typically have a bandwidth constraint. In this case, the problem of evaluating performance of a unipolar OFDM scheme is nontrivial; see \cite{MCEA,ZZ17,TSH121}.
For a unipolar OFDM block $\textbf{s}$, the corresponding waveform generated by a digital-to-analog converter (DAC) is
\begin{equation}
\label{waveform}
s(t)=\sum_{n=0}^{N-1}s_n g\left(t-nT\right)+{\cal A}, \mspace{8mu}s(t)\ge 0
\end{equation}
where $g(t)$ is a shaping pulse. Note that pulse shaping often destroys nonnegativity \cite{TSH121,TSH13,ZZ17,MCEA} and a DC bias is then needed.
We can study the information rates of unipolar OFDM schemes in bandlimited Gaussian OICs using the bounding technique in \cite{ZZ17}. Optimizing $g(t)$ to boost the information rate of $s(t)$ is an interesting problem with practical importance.

\begin{appendices}
      \section{A General Framework for Transmission over Complex-Valued Vector Channels with Transceiver Distortion}

In this appendix we introduce the general framework for transmission with transceiver distortion in \cite{Boss}, and extend it to complex-valued vector channels.

Consider a complex-valued vector channel with input $\textbf{X}^{(i)}$ $=\left[X^{(i)}_{1},...,X^{(i)}_{N}\right]^\textrm T$ and additive noise $\textbf{Z}^{(i)}=\left[Z^{(i)}_{1},...,Z^{(i)}_{N}\right]^\textrm T$ where $i$ is the time index, $\textbf{Z}^{(i)}$ is ergodic in terms of blocks and is independent of $\textbf{X}^{(i)}$. The channel output is a deterministic mapping $f(\cdot)$ which transforms a pair of channel input vector and noise vector $\left\{\textbf{X}^{(i)},\textbf{Z}^{(i)}\right\}$ into a vector
\begin{equation}
\label{VecC}
\textbf{Y}^{(i)}=f\left(\textbf{X}^{(i)},\textbf{Z}^{(i)}\right)=f_\textrm O\left(f_\textrm I\left(\textbf{X}^{(i)}\right)+\textbf{Z}^{(i)}\right),
\end{equation}
where $i=1,...,\ell,$, $f_\textrm I$ and $f_\textrm O$ are the mappings in the transmitter and the receiver, respectively, $\ell$ denotes the codeword length in terms of vectors. Eqn. (\ref{VecC}) is a general model of channels with nonlinear transceiver distortions which are memoryless in terms of vectors or blocks, e.g., nonlinear transfer characteristics of transmitter,
I/Q imbalances, the analog-to-digital conversion (i.e., quantization) at receiver, OFDM with clipping, and so on.

For transmission of rate $\cal R$ nats per block, assume that a message $m$ is selected from ${\cal M} =\left\{1,...,\lfloor e^{\ell R}\rfloor\right\}$ uniformly randomly. The encoder maps $m$ to a length-$\ell N$ transmitted codeword
$\left\{\textbf{\texttt{X}}^{(i)}(m)\right\}_{i=1}^{\ell}$
in an ICG codebook,
where $\textbf{\texttt{X}}^{(i)}(m)=\left[\texttt{X}^{(i)}_{1}(m),...,\texttt{X}^{(i)}_{N}(m)\right]^\textrm T$, $1\leq i\leq\ell$.
The codebook is generated from an ensemble of ICG codebooks. That is, all the codewords are generated with each element i.i.d. according to a complex Gaussian distribution.
At the receiver, we let the decoder follow a nearest neighbor decoding rule as
$\hat{m}=\mathop{\arg\min}_{\textsf{m}\in{\cal M}}\textsf{d}_\textrm E(\textsf{m})$
where
\begin{equation}
\label{DA}
\textsf{d}_\textrm E(\textsf{m})=\frac{1}{\ell}\sum_{i=1}^{\ell}\left\|\textbf{Y}^{(i)}-a\textbf{\texttt{X}}^{(i)}(\textsf{m})\right\|^2,\mspace{8mu} \textsf{m}\in {\cal M}, \mspace{8mu} a\in \mathbb{C},
\end{equation}
i.e., in the codebook, the decoder finds the codeword that minimizes its (scaled) Euclidean distance to the received vector.
In (\ref{DA}), $a$ is a decoding scaling parameter to be optimized.

For any transmission rate $\cal R$ below the GMI, the average probability of decoding error of the decoder described above (averaged over both the messages and the codebook ensemble) decreases to zero as the channel coding length grows without bound \cite{GLT}, \cite{Boss}. Following essentially the same steps as [\ref{Boss}, Appendix C], the GMI of the channel (\ref{VecC}) achieved by an ICG codebook ensemble and the decoding metric (\ref{DA}) can be obtained as
\begin{equation}
\label{GMI}
{\cal I}_\textrm{GMI}=N\log\left(1+\frac{\Delta}{1-\Delta}\right),
\end{equation}
where
\begin{equation}
\Delta=\frac{\left(\textrm E\left[({f}(\textbf{X},\textbf{Z}))^\textrm H \textbf{X}\right]\right)^2}{\textrm E\left[\|\textbf{X}\|^2\right]\textrm E\left[\|f(\textbf{X},\textbf{Z})\|^2\right]},
\end{equation}
and the expectation operation $\textrm E\left[\cdot\right]$ is taken with respect to $\textbf{X}$ and $\textbf{Z}$. Here, the scaling parameter $a$ should be set as
\begin{equation}
\label{aoptA}
a_\textrm{opt}=\frac{\textrm E\left[(f(\textbf{X},\textbf{Z}))^\textrm T \bar{\textbf{X}}\right]}{\textrm E\left[\|\textbf{{X}}\|^2\right]}
\end{equation}
to achieve ${\cal I}_\textrm{GMI}$ given in (\ref{GMI}).
In this paper, the distortion occurs only at the transmitter. So we have
\begin{equation}
\textbf{Y}=f_\textrm I(\textbf{X})+\textbf{Z},
\end{equation}
\begin{equation}
\Delta=\frac{\left(\textrm E\left[(f_\textrm I(\textbf{X}))^\textrm H \textbf{X}\right]\right)^2}{\textrm E\left[\|\textbf{X}\|^2\right]\left(\textrm E\left[\|f_\textrm I(\textbf{X})\|^2\right]+\textrm E \left[\|\textbf{Z}\|^2\right]\right)},
\end{equation}
and
\begin{equation}
a_\textrm{opt}=\frac{\textrm E\left[(f_\textrm I(\textbf{X}))^\textrm T \bar{\textbf{X}}\right]}{\textrm E\left[\|\textbf{{X}}\|^2\right]}.
\end{equation}

In addition, for any transmission rate $\cal R$ above the GMI, the average probability of decoding error of the decoder described above, further averaged over the ensemble of i.i.d. Gaussian codebook ensembles, tends to one as the channel coding length grows without bound \cite{GLT,LS}. So the GMI can be interpreted as the maximally achievable information rate of a ``typical'' codebook which is generated without regard to the specific operations in unipolar OFDM schemes such as clipping and nearest neighbor decoding.

\end{appendices}

\ifCLASSOPTIONcaptionsoff
  \newpage
\fi


\begin{thebibliography}{1}
\bibitem{WCSP}
J. Zhou and W. Zhang, ``Information rates of unipolar OFDM schemes in Gaussian optical intensity channel,'' in \emph{Proc. Ninth Int. Conf. Wirel. Commun. Signal Process. (WCSP)}, Nanjing, China, Oct. 2017.

\bibitem{Armstrong}
J. Armstrong, ``OFDM for optical communications,'' \emph{J. Lightw. Technol.}, vol. 27, no. 3, pp. 189--204, Feb. 2009.

\bibitem{B94}
\label{B94}
J. R. Barry, \emph{Wireless Infrared Communications}. Boston: Kluwer, 1994.

\bibitem{CK96}
J. B. Carruthers and J. M. Kahn, ``Multiple-subcarrier modulation for nondirected wireless infrared communication,'' \emph{IEEE J. Sel. Areas
Commun.}, vol. 14, no. 3, pp. 538--546, Apr. 1996.

\bibitem{AL06}
\label{AL06}
J. Armstrong and A. J. Lowery, ``Power efficient optical OFDM,'' \emph{Electron. Lett.}, vol. 42, no.6, pp. 371--372, Mar. 2006.

\bibitem{LRBK09}
\label{LRBK09}
S. C. J. Lee, S. Randel, F. Breyer, and A. M. J. Koonen, ``PAM-DMT for intensity-modulated and direct-detection optical communication systems,'' \emph{IEEE Photon. Technol. Lett.}, vol. 21, no. 23, pp. 1749--1751, Dec. 2009.

\bibitem{Yong07}
J. Yong, ``Modulation and demodulation apparatuses and methods for wired/wireless communication,'' Korea Patent WO2007/064 165 A, 07,
2007.

\bibitem{FHV11}
N. Fernando, Y. Hong, and E. Viterbo, ``Flip-OFDM for unipolar communication systems,'' \emph{IEEE Trans. Commun.}, vol. 60, no. 12, pp. 3726--3733, Dec. 2012.

\bibitem{TSH12}
D. Tsonev, S. Sinanovic, and H. Haas, ``Novel unipolar orthogonal frequency division multiplexing (U-OFDM) for Optical Wireless,'' in \emph{Proc. IEEE Vehic. Tech. Conf. (VTC Spring)}, Yokohama, Japan, May 2012.

\bibitem{NGHL12}
A. Nuwanpriya, A. Grant, S.-W. Ho, and L. Luo, ``Position modulating OFDM for optical wireless communications,'' in \emph{Proc. 3rd IEEE Globecom Workshop Opt. Wirel. Commun. (OWC)}, Anaheim, CA, USA, Dec. 2012, pp. 1219--1223.

\bibitem{DPA11}
S. D. Dissanayake, K. Panta, and J. Armstrong, ``A novel technique to simultaneously transmit ACO-OFDM and DCO-OFDM in IM/DD systems,'' in \emph{ Proc. 2nd IEEE Globecom Workshop Opt. Wirel. Commun. (OWC)}, Houston, TX, USA, Dec. 2011, pp. 782--786.

\bibitem{RK14}
B. Ranjha and M. Kavehrad, ``Hybrid asymmetrically clipped OFDM-based IM/DD optical wireless system,'' \emph{IEEE/OSA J. Opt. Commun. Netw.}, vol. 6, no. 4, pp. 387--396, Apr. 2014.

\bibitem{WB15}
N. Wu and Y. Bar-Ness, ``A novel power-efficient scheme asymmetrically and symmetrically clipping optical (ASCO)-OFDM for IM/DD optical systems,'' \emph{EURASIP J. Adv. Signal Process.}, vol. 2015, no. 3, Jan. 2015.

\bibitem{EL14}
H. Elgala and T. D. C. Little, ``SEE-OFDM: Spectral and energy efficient OFDM for optical IM/DD systems,'' in \emph{Proc. IEEE 25th Ann. Int. Symp. Personal Indoor Mob. Radio Commun. (PIMRC)}, Washington DC, USA, Sept. 2014, pp. 851--855.

\bibitem{LWEL}
E. Lam, S. K. Wilson, H. Elgala, and T. D. C. Little, ``Spectrally and energy efficient OFDM (SEE-OFDM) for intensity modulated optical wireless systems,'' \emph{arXiv preprint} arXiv:1510.08172, 2015.

\bibitem{W15}
Q. Wang, C. Qian, X. Guo, Z. Wang, D. G. Cunningham, and I. H. White, ``Layered ACO-OFDM for intensity-modulated direct-detection optical wireless transmission,'' \emph{Opt. Express}, vol. 23, no. 9, pp. 12382--12393, May 2015.

\bibitem{ITH1}
M. S. Islim, D. Tsonev, and H. Haas, ``On the superposition modulation for OFDM-based optical wireless communication,'' in \emph{Proc. 3rd IEEE Global Conf. Signal Inf. Process. (GlobalSIP),} Orlando, FL, USA, Dec. 2015, pp. 1022--1026.

\bibitem{TVH15}
D. Tsonev, S. Videv, and H. Haas, ``Unlocking spectral efficiency in intensity modulation and direct detection systems,'' \emph{IEEE J. Sel. Areas
Commun.}, vol. 33, no. 9, pp. 1758--1770, Sept. 2015.

\bibitem{Li1}
X. Li, R. Mardling, and J. Armstrong, ``Channel capacity of IM/DD optical communication systems and of ACO-OFDM,'' in \emph{Proc.
IEEE 2007 Int. Conf. Commun. (ICC)}, Glasgow, U.K., Jun. 2007, pp. 2128--2133.

\bibitem{ITH2}
M. S. Islim, D. Tsonev, and H. Haas, ``Spectrally enhanced PAM-DMT for IM/DD optical wireless communications,'' in \emph{Proc. IEEE 26th Ann. Int. Symp. Personal Indoor Mob. Radio Commun. (PIMRC)}, Hongkong, China, Aug.--Sept. 2015, pp. 877--882.

\bibitem{MChen}
R. Guan, N. Huang, H. Wang, J.-Y. Wang, and M. Chen, ``Novel spectral efficient OFDM for optical wireless communication'', in \emph{Proc. IEEE 2016 Int. Conf. Commun. (ICC)}, Kuala Lumpur, Malaysia, May. 2016.

\bibitem{EL13}
H. Elgala and T. D. C. Little, ``Reverse polarity optical OFDM (RPO-OFDM): dimming compatible OFDM for gigabit VLC links,'' \emph{Opt. Express}, vol. 21, no. 20, pp. 24288--24299, Oct. 2013.

\bibitem{WWD15}
Q. Wang, Z. Wang, and L. Dai, ``Asymmetrical hybrid optical OFDM for visible light communications with dimming control,'' \emph{IEEE Photon. Technol. Lett.}, vol. 27, no. 9, pp. 974--977, May 2015.

\bibitem{YGL16}
F. Yang, J. Gao, and S. Liu, ``Novel visible light communication approach based on hybrid OOK and ACO-OFDM,'' \emph{IEEE Photon. Technol. Lett.}, vol. 28, no. 14, pp. 1585--1588, Jul. 2016.

\bibitem{AFH11}
K. Asadzadeh, A. A. Farid, and S. Hranilovic, ``Spectrally factorized optical OFDM,'' in \emph{Proc. IEEE 12th Can. Workshop Inf. Theory (CWIT)}, Kelowna, BC, Candada, May 2011, pp. 102--105.

\bibitem{MMJ10}
M. S. Moreolo, R. Munoz, and G. Junyent, ``Novel power efficient optical OFDM based on Hartley transform for intensity-modulated direct-detection systems,'' \emph{J. Lightw. Technol.}, vol. 28, no. 5, pp. 798--805, Mar. 2010.

\bibitem{Zhou2}
J. Zhou, Y. Qiao, T. Zhang, E. Sun, M. Guo, Z. Zhang, X. Tang, and F. Xu, ``FOFDM based on discrete cosine transform for intensity-modulated and direct-detected systems'', \emph{J. Lightw. Technol.}, vol. 34, no. 16, pp. 3717--3725, Aug. 2016.

\bibitem{EL141}
H. Elgala and T. D. Little, ``P-OFDM: spectrally efficient unipolar OFDM,'' in \emph{Proc. Opt. Fiber Commun. Conf. (OFC)}, San Francisco, CA, USA, Aug. 2014.

\bibitem{SEUYou}
J. Xu, W. Xu, H. Zhang, and X. You, ``Asymmetrically reconstructed optical OFDM for visible light communications,'' \emph{IEEE Photon. J.}, vol. 8., no. 1, Feb. 2016.

\bibitem{ZCZhang}
L. Wu, Z. Zhang, J. Dang, J. Wang, and H. Liu, ``Polarity information coded flip-OFDM for intensity modulated systems,'' \emph{IEEE Commmun. Lett.}, vol. 20, no. 8, pp. 1535--1538, Aug. 2016.

\bibitem{BWW17}
R. Bai, Q. Wang, and Z. Wang, ``Asymmetrically clipped absolute value optical OFDM for intensity-modulated direct-detection systems,'' \emph{J. Lightw. Technol.}, vol. 35, no. 17, pp. 3680--3691, Sept. 2017.

\bibitem{Xu}
Q. Gao, C. Gong, S. Li, and Z. Xu, ``DC-informative modulation for visible light communications under lighting constraints,'' \emph{IEEE Wirel. Commun.}, vol. 22, no. 2, pp. 54--60, Apr. 2015.

\bibitem{YZFG}
Y. Yang, Z. Zeng, S. Feng, and C. Guo, ``A simple OFDM scheme for VLC systems based on $\mu$-law mapping.'' \emph{IEEE Photon. Technol. Lett.}, vol. 28, no. 6, pp. 641--644, Mar. 2016.

\bibitem{BUPT}
N. Yin, C. Guo, Y. Yang, P. Luo, and C. Feng, ``Asymmetrical and direct current biased optical OFDM for visible light communication with dimming control,'' in \emph {Proc. 3rd IEEE ICC Workshop Opt. Wirel. Commun. (OWC)}, Paris, France, May 2017.

\bibitem{JS09}
J. Armstrong and B. J. C. Schmidt, ``Comparison of asymmetrically clipped optical OFDM and DC-biased optical OFDM in AWGN,'' \emph{IEEE
Commun. Lett.}, vol. 12, pp. 343--345, 2008.

\bibitem{MEH11}
R. Mesleh, H. Elgala, and H. Haas, ``On the performance of different OFDM based optical wireless communication systems,'' \emph{IEEE/OSA J. Opt. Commun. Netw.}, vol. 3, no. 8, pp. 620--628, Aug. 2011.

\bibitem{DA13}
S. D. Dissanayake and J. Armstrong, ``Comparison of ACO-OFDM, DCO-OFDM and ADO-OFDM in IM/DD systems,'' \emph{J. Lightw. Technol.}, vol. 31, no. 7, pp. 1063--1072, 2013.

\bibitem{TSH13}
D. Tsonev, S. Sinanovic, and H. Haas, ``Complete Modeling of Nolinear Distortion in OFDM-Base Optical Wireless Communication,'' \emph{J. Lightw. Technol.}, vol. 31, no. 18, pp. 3064--3076, Sept. 2013.

\bibitem{Lowery}
A. J. Lowery, ``Comparisons of spectrally-enhanced asymmetrically-clipped optical OFDM systems,'' \emph{Opt. Express}, vol. 24, no. 4, pp. 3950--3966, Feb. 2016.

\bibitem{SYG17}
Y. Sun, F. Yang, and J. Gao, ``Comparison of hybrid optical modulation schemes for visible light communication,'' \emph{IEEE Photon. J.}, vol. 9, no. 3, Jun. 2017.

\bibitem{Li2}
X. Li, J. Vucic, V. Jungnickel, and J. Armstrong, ``On the capacity of intensity-modulated direct-detection systems and the information rate
of ACO-OFDM for indoor optical wireless applications,'' \emph{IEEE Trans. Commun.}, vol. 60, no. 3, pp. 799--809, Mar. 2012.

\bibitem{Yu1}
Z. Yu, R. J. Baxley, and G. T. Zhou, ``EVM and achievable data rate analysis of clipped OFDM signals in visible light communication,'' \emph{EURASIP J. Wireless Commun. Netw.}, vol. 2012, Oct. 2012.

\bibitem{Yu2}
Z. Yu, R. J. Baxley, and G. T. Zhou, ``Achievable data rate analysis of clipped Flip-OFDM in optical wireless communication,'' in \emph{Proc. 3rd IEEE Globecom Workshop Opt. Wirel. Commun. (OWC)}, Anaheim, CA, USA, Dec. 2012, pp. 1203--1207.

\bibitem{DH13}

S. Dimitrov and H. Haas, ``Information rate of OFDM-based optical wireless communication systems with nonlinear distortion,'' \emph{J. Lightw. Technol.}, vol. 31, no. 6, pp. 918--929, Mar. 2013.

\bibitem{HK04}
\label{bHK04}
S. Hranilovic and F. R. Kschischang, ``Capacity bounds for power- and band-limited optical intensity channels corrupted by Gaussian noise,'' \emph{IEEE Trans. Inf. Theory}, vol. 50, no. 5, pp. 784--795, May 2004.

\bibitem{LMW09}
\label{LMW09}
A. Lapidoth, S. M. Moser, and M. A. Wigger, ``On the capacity of free-space optical intensity channels,'' \emph{IEEE Trans. Inf. Theory}, vol. 55, no. 10, pp. 4449--4461, Oct. 2009.

\bibitem{FH09}
\label{FH09}
A. A. Farid and S. Hranilovic, ``Channel capacity and non-uniform signalling for free-space optical intensity channels,'' \emph{IEEE J. Sel. Areas Commun.}, vol. 27, no. 9, pp. 1553--1563, Dec. 2009.

\bibitem{FH10}
\label{FH10}
A. A. Farid and S. Hranilovic, ``Capacity bounds for wireless optical intensity channels with Gaussian noise,'' \emph{IEEE Trans. Inf. Theory}, vol. 56, no. 12, pp. 6066--6077, Dec. 2010.

\bibitem{WHWCW}
J.-B. Wang, Q.-S. Hu, J. Wang, M. Chen, and J-.Y-. Wang, ``Tight bounds on channel capacity for dimmable visible light communications,'' \emph{J. Lightw. Technol.}, vol. 31, no. 23, pp. 3771--3779, Dec. 2013.

\bibitem{CMA15}
\label{bCMA15}
A. Chaaban, J. Morvan, and M.-S. Alouini, ``Free-space optical communications: capacity bounds, approximations, and a new sphere packing perspective,'' \emph{IEEE Trans. Commun.}, vol. 64, no. 3, pp. 1176--1191, Mar. 2016.

\bibitem{ZZ17}
J. Zhou and W. Zhang, ``On the capacity of bandlimited optical intensity channels with Gaussian noise,'' \emph{IEEE Trans. Commun.}, vol. 65, no. 6, pp. 2481--2493, Jun. 2017.

\bibitem{FU98}

G. D. Forney and G. Ungerboeck, ``Modulation and coding for linear Gaussian channels,'' \emph{IEEE Trans. Inf. Theory}, vol. 44, no.6, pp. 2384--2415, Oct. 1998.

\bibitem{GLT}
A. Ganti, A. Lapidoth, and \.I. E. Telatar, ``Mismatched decoding revisited: general alphabets, channels with memory, and the wide-band limit,'' \emph{IEEE Trans. Inf. Theory}, vol. 46, no. 7, pp. 2315--2328, Nov. 2000.

\bibitem{LS}
A. Lapidoth and S. Shamai (Shitz), ``Fading channels: how perfect need `perfect side information' be?'' \emph{IEEE Trans. Inf. Theory}, vol. 48, no. 5, pp. 1118--1134, May 2002.

\bibitem{Boss}
\label{Boss}
W. Zhang, ``A general framework for transmission with transceiver distortion and some applications,'' \emph{IEEE Trans. Commun.}, vol. 60, no. 2, pp. 384--399, Feb. 2012.

\bibitem{Smith71}
\label{b14}
J. G. Smith, ``The information capacity of amplitude- and variance constrained scalar Gaussian channels," \emph{Inf. Contr.}, vol. 18, no. 3, pp. 203--219, Feb. 1971.

\bibitem{DSH121}
S. Dimitrov, S. Sinanovic, and H. Haas, ``Signal shaping and modulation for optical wireless communication,'' \emph{J. Lightw. Technol.}, vol. 30, no. 9, pp. 1319--1328, May 2012.

\bibitem{Wu}
N. Wu and Y. Bar-Ness, ``Lower bounds on the channel capacity of ASCO-OFDM and ADO-OFDM,'' in \emph{Proc. 49th Ann. Conf. Inf. Sci. Syst. (CISS)}, Baltimore, MD, USA, Mar. 2015.

\bibitem{Bussgang}
J. J. Bussgang, ``Crosscorrelation functions of amplitude-distorted Gaussian signals,'' Technical Report No. 216, Research Laboratory of Electronics, Massachusetts Institute of Technology, Cambridge, MA, Mar. 1952.

\bibitem{OI}
H. Ochiai and H. Imai, ``Performance analysis of deliberately clipped OFDM signals,'' \emph{IEEE Trans. Commun.}, vol. 50, no. 1, pp. 89--101, Jan. 2002.

\bibitem{HH}
B. Hassibi and B. M. Hochwald, ``How much training is needed in multiple-antenna wireless links?'' \emph{IEEE Trans. Inf. Theory}, vol. 49, no. 4, pp. 951--963. Apr. 2003.

\bibitem{Lapidoth}
A. Lapidoth, ``Nearest neighbor decoding for additive non-Gaussian noise channels,'' \emph{IEEE Trans. Inf. Theory}, vol. 42, no. 5, pp. 1520--1529, Sept. 1996.

\bibitem{SEUZhang}
M. Zhang and Z. Zhang, ``An optimum DC-biasing for DCO-OFDM system,'' \emph{IEEE Commun. Lett.}, vol. 18, no. 8, pp. 1351--1354, Aug. 2014.

\bibitem{SEUZhao}
X. Ling, J. Wang, X. Liang, Z. Ding, and C. Zhao, ``Offset and power optimization for DCO-OFDM in visible light communication systems,'' \emph{IEEE Trans. Signal Process.}, vol. 64, no. 2, pp. 349--363, Jan. 2016.

\bibitem{TSH121}
D. Tsonev, S. Sinanovic, and H. Haas, ``Pulse shaping in unipolar OFDM-based modulation schemes,'' in \emph{Proc. 3rd IEEE Globecom Workshop Opt. Wirel. Commun. (OWC)}, Anaheim, CA, USA, Dec. 2012, pp. 1208--1212.

\bibitem{MCEA}
S. Mazahir, A. Chaaban, H. Elgala, and M.-S. Alouini, ``Effective information rates of single-carrier and multi-carrier modulation schemes for bandwidth constrained IM/DD systems,'' in \emph{Proc. IEEE 2017 Int. Conf. Commun. (ICC)}, Paris, France, May 2017.

\end{thebibliography}
 \end{document}